\documentclass[twocolumn]{aastex62}

\usepackage{natbib}
\usepackage{subfigure}
\usepackage{url}
\usepackage{amsmath}
\usepackage[normalem]{ulem}
\usepackage{bm}
\usepackage{comment}
\usepackage{array}
\newcolumntype{P}[1]{>{\centering\arraybackslash}p{#1}}
\newcolumntype{M}[1]{>{\centering\arraybackslash}m{#1}}
\usepackage{color}
\usepackage{soul}

\newcommand{\kep}{{\it Kepler}}
\newcommand{\kt}{{\it K2}}

\newcommand{\rearth}{{R$_\oplus$}}
\newcommand{\rsun}{{R$_\odot$}}
\newcommand{\msun}{{M$_\odot$}}
\newcommand{\lsun}{{L$_\odot$}}

\newcommand{\logg}{{log(g)}}

\newcommand{\feh}{{[Fe/H]}~}

\newcommand{\thisstar}{{K2-146}}

\newcommand{\washington}{Department of Astronomy, University of Washington, Box 351580, 3910 15th Ave NE, Seattle, WA 98195, USA}
\newcommand{\chicago}{Department of Astronomy and Astrophysics, University of
Chicago, 5640 S. Ellis Ave, Chicago, IL 60637, USA}
\newcommand{\sagan}{Sagan Fellow}
\newcommand{\goddard}{NASA Goddard Space Flight Center, Greenbelt, MD, 20771, USA}

\begin{document}

\title{K2-146: Discovery of Planet c, Precise Masses from Transit Timing, and Observed Precession}

\shorttitle{K2-146} 
\shortauthors{Hamann et al.}

\author[0000-0003-3996-263X]{Aaron~Hamann}
\affiliation{\chicago}

\author[0000-0001-7516-8308]{Benjamin~T.~Montet}
\altaffiliation{\sagan}
\affiliation{\chicago}

\author[0000-0003-3750-0183]{Daniel~C.~Fabrycky}
\affiliation{\chicago}

\author[0000-0002-0802-9145]{Eric~Agol}
\affiliation{\washington}

\author[0000-0002-0493-1342]{Ethan~Kruse}
\affiliation{\goddard}

\correspondingauthor{Aaron Hamann}
\email{ahamann@uchicago.edu}

\begin{abstract}
K2-146 is a mid-M dwarf ($M_\star = 0.331 \pm 0.009 M_\odot$; $R_\star = 0.330 \pm 0.010 R_\odot$), observed in Campaigns 5, 16, and 18 of the \kt\ mission. In Campaign 5 data, a \added{single} planet was discovered with an orbital period of $2.6$~days and large transit timing variations due to an unknown perturber. Here we analyze data from Campaigns 16 and 18, detecting the transits of a \replaced{planet c}{second planet, c,} with an orbital period of $4.0$~days, librating in a 3:2 resonance with planet b. Large, anti-correlated timing variations of both planets exist due to their resonant perturbations. The planets have a mutual inclination of $2.40^\circ\pm0.25^\circ$, which torqued planet c more closely into our line-of-sight. Planet c was grazing in Campaign 5 and thus missed in previous searches; in Campaigns 16 and 18 it is fully transiting, and its transit depth is three times larger. We improve the stellar properties using data from Gaia DR2, and \replaced{use our}{using} dynamical fits \deleted{to} find that both planets are sub-Neptunes: their masses are $5.77\pm0.18$ and $7.50\pm0.23 M_{\oplus}$ and their radii are $2.04\pm0.06$ and $2.19\pm0.07$ \rearth, respectively. These mass constraints set the precision record for small exoplanets (a few gas giants have comparable relative precision). These planets lie in the photoevaporation valley when viewed in Radius-Period space, but due to the low-luminosity M-dwarf host star, they lie among the atmosphere-bearing planets when viewed in Radius-Irradiation space. This, along with their densities being 60\%-80\% that of Earth, suggests that they may both have retained a substantial gaseous envelope.
\end{abstract}


\section{Introduction} \label{sec:intro}

The \kt\ mission \citep{2014Howell} observed 20 fields across the ecliptic plane. Because of the limitations of the spacecraft, with the loss of two reaction wheels, the telescope could only stably point in limited directions.
With much of the ecliptic plane already observed by the telescope, later campaign fields overlapped with previous fields and some stars were observed in two or three campaigns.

Campaign 5 was observed as a part of the \kt\ mission from 2015 Apr 27 to 2015 Jul 10. Campaign 16 revisited a fraction of this field from 2017 Dec 07 to 2018 Feb 25, while Campaign 18 again revisited this field from 2018 May 12 to 2018 Jul 02. These observations combine to provide 205 days of observations over the ``\kt\ Legacy Field'' spread over 3.2 years. By combining these data for stars observed by multiple campaigns, planetary systems can be characterized in higher detail than for each individual campaign alone \citep[e.g.][]{2018Chakraborty}.
Moreover, when planets orbiting the same star are dynamically interacting, 
the long observing baseline provides an opportunity to observe and understand transit timing variations (TTVs) to measure planet masses and orbital parameters, as was commonly done with data from the original \kep\ mission \citep[e.g.][]{2013Huber, 2014Hadden, 2016JontofHutter}.

\subsection{K2-146 system}

One of the stars observed in Campaigns 5, 16, and 18 was EPIC 211924657, also named K2-146 after the validation of its first planet.\footnote{This star was proposed as a target in many programs. For Campaign 5: GO5020, GO5097, GO5054, GO5011, and GO5006. For Campaign 16: GO16005, GO16011, GO16052, GO16077, GO16101, and GO16015, the final two of which requested short cadence observations. For Campaign 18: GO18068, GO18027, GO18048, and GO18063, of which all but the first requested short cadence observations.} When Campaign 5 data was available, \citet{2016Pope} quickly identified a planet candidate with a 2.6 day period; the planet was also detected by the \citet{2018Petigura} search, and the host star and planet better characterized with follow-up spectroscopy by \citet{2017Dressing} --- who also noted that the planet showed TTVs. \citet{2018Hirano} performed a detailed analysis of the system, validating the planet by combining spectroscopy and adaptive optics imaging with the \kt\ data. These authors also noticed the presence of TTVs, finding a rapid change in the observed orbital period of $\sim 10$~minutes, at time BKJD\footnote{BKJD = BJD - 2454833.}$=2360$. \citet{2018Livingston} similarly analyzed and validated planet b while noting the existence of TTVs. Each of these teams only reported the presence of a single transiting planet in Campaign 5, and hence were not able to interpret the cause of these TTVs.

\subsection{Precessing Planets}

It was recognized early in the history of transiting exoplanets that a planet with a large impact parameter, nearly grazing its star, is exquisitely sensitive to additional planets via the out-of-plane torque exerted between the orbits \citep{2002MiraldaEscude}. A changing inclination would generate large transit duration/depth variations (TDVs). In a first example, tentative upper limits on the transit depth of GJ436b seemed to conflict with later measurements of its transits, indicating that its inclination was changing on yearly timescales \cite{2008Ribas}. Later, a series of transits of GJ436b were shown to have a constant shape \citep{2010Ballard}, providing upper limits on inclination change. Thus a limit was placed on the product of the mass, $m_c$, and rotation around the line-of-sight, $\Delta \Omega$, of secular perturbers of various periods. Our detection, to be described, allows us to constrain planetary masses by TTVs and then determine $\Delta \Omega$ via interpreting the depth change, much as envisioned by \cite{2008Ribas}. 

Since then, several more robust examples of orbital precession in exoplanets have been discovered.

First, Kepler-13 was shown to have both a non-symmetric transit shape and a changing transit duration \citep{2012Szabo}. Both are attributed to a misaligned orbit around a rapidly spinning host star: the former is due to the temperature decrease from pole to equator, and the latter is due to the torque from the stellar oblateness.

Second, several circumbinary planets have orbits misaligned from the orbital plane of the host binary; the torque from the binary causes the planets' orbits to precess on decade-long timescales \citep{2012Welsh, 2014Kostov}. 

Finally, three planetary systems have shown planet-planet interactions that precess one another's orbits, Kepler-117 \citep{2015Almenara}, Kepler-9 \citep{2018Freudenthal,2019Borsato} and Kepler-119 \citep{2017MillsFabrycky}. The latter requires a large mutual inclination of $24^{+11}_{-8}$~degrees to fit the TTVs and TDVs without invoking additional planets. 

Another effect that can be probed by TDVs is in-plane precession of eccentricity. The resonant term of periastron torque (on the libration/TTV period) has been inferred by \cite{2013Nesvorny}, but so far no claim of a secular term has been made, to our knowledge. As described by \cite{2008Pal}, (a) at low impact parameter, precession of the periastron affects TDVs by changing the velocity of the planet at transit: transit near periastron yields a shorter than average duration; (b) at high impact parameters, precession changes the distance between the planet and the star at transit, which affects the transit chord length: transit near periastron yields a longer transit chord and a longer than average duration. For small orbital eccentricities, the impact parameter separating these two regimes is $\sqrt{1/2}$, and we shall show that the two-planet system K2-146 furnishes one example in each of these regimes. These sources of TDVs, and others, are described in \citet{Agol2018}.

\subsection{Verification of TTV perturbers}

TTVs were proposed as a method of detecting companions of transiting exoplanets \citep{Agol05, Holman1288}. A clear precedent for our discovery has been made in Kepler-36 \citep{2012Carter}, where a planet with a 16-day orbital period was seen to exhibit strong and sharp (non-sinuosidal) TTVs. It took about two years of data to recognize that its perturber was also transiting, after which it was immediately clear that the new, 13-day planet could explain the TTV pattern. 

Recently there have been a few other examples of systems where a single planet has TTVs, and the perturber was predicted and later verified by radial velocity. \cite{2014Barros} were the first to achieve this, finding the perturbing companion predicted by \cite{2013Nesvorny}. A planetary perturber was inferred in the Kepler-19 system by \cite{2011Ballard}, albeit with incomplete information, and a radial-velocity planet with the right properties was later detected by \cite{2017Malavolta}. In the Kepler-419 system, a highly eccentric planet's TTVs predicted an additional planet \citep{2012Dawson} which was then confirmed and further characterized by radial velocity measurements \citep{2014Dawson}. A perturbing planet's mass and period was determined via highly precise TTV data by \cite{2012Nesvorny}; \deleted{though it is completely convincing as a bona fide detection,} the second planet \added{candidate} has not yet been detected by other means. In none of these systems is the perturbing planet known to transit, but with enough mutual inclination, it might be torqued by the currently-transiting planet into a transiting orbit of its own. 
\qquad

In Section~\ref{sec:analysis} we discuss how we update the understanding of K2-146 using Gaia data, how we treat the light curve, and how we model the transit timing. Section~\ref{sec:results} gives the numerical results. In Section~\ref{sec:dynamics} we discuss the dynamical evolution of the system. Finally, in Section~\ref{sec:discussion} we conclude with inferences we may draw and further steps worth taking in the study of this exciting system. 

\section{Data Analysis}
\label{sec:analysis}

\subsection{Stellar Parameters enabled by Gaia}
\label{sec:gaia}

\citet{2018Hirano} obtained a spectrum of \thisstar\ with the
High Dispersion Spectrograph on the Subaru 8.2 m telescope \citep{2002Noguchi}.
\replaced{From a fit to this spectrum, those authors measured}{Those authors inferred} a stellar effective temperature
of $3385 \pm 70$ K, a metallicity \feh of $-0.02 \pm 0.12$, and a surface gravity
\logg\ of $4.906 \pm 0.041$. 
From these stellar parameters, the authors infer a stellar radius of $0.350 \pm 0.035$ \rsun.
This star was also observed by \citet{2018Rodriguez}, who use near-IR spectroscopy from Palomar/TripleSpec to infer a temperature of $3766 \pm 195$ K, a log luminosity of $-1.91 \pm 0.16$ \lsun, and a radius of $0.68 \pm 0.16$ \rsun.
\added{While these uncertainties are large, the point estimates are considerably larger than those from \citet{2018Hirano}. From the published spectra in \citet{2018Rodriguez}, the star was observed at a lower signal-to-noise ratio than the typical star in that survey, likely leading to the inflated uncertainties on inferred stellar parameters. This star was also observed spectroscopically as a part of the LAMOST survey \citep{2012Zhao}, with parameters described in the fourth data release. The LAMOST pipeline identifies this target as an M3 dwarf, consistent with the parameters of \citet{2018Hirano}.}

We combine the spectroscopic stellar parameters \added{from \citet{2018Hirano}} with data from Gaia DR2 \citep{2016Gaia, 2018Gaia} in order to improve the precision on the inferred radius.
The published DR2 parallax for this star is $12.5821 \pm 0.0750$ milliarcsec.
We apply a correction of $+0.029$ mas to this value to account for the zeropoint
offset inherent in the DR2 results \citep{2018Lindegren}. We use this parallax, the 
inferred stellar parameters from \citet{2018Hirano}, and JHK photometry from the
2-Micron All Sky Survey \citep[2MASS,][]{2003Cutri}, as listed in Table 1, as inputs to the \texttt{isochrones} package, a Python interface to fit derived stellar parameters, photometry, and parallax information simultaneously to stellar models \citep{2015Morton}. 

Fitting these data to the MESA Isochrones and Stellar Tracks (MIST) models \citep{2016Choi}, we find an inferred radius of $0.323 \pm 0.008$ \rsun\ and mass of $0.323 \pm 0.011$ \msun.
Alternatively, fitting to the Dartmouth Stellar Evolution Database \citep{2008Dotter} result 
in a radius of $0.330 \pm 0.005$ \rsun\ and mass of $0.331 \pm 0.007$ \msun.
We choose to use the results of the fit to the Dartmouth models, which provide a higher maximum log-likelihood fit to our input data $\Delta \mathcal{\log L} = 2.92$, 
although we note that both sets of stellar models may suffer from systematic uncertainties at the few percent level in their inference of stellar radii.
\citet{2015Mann} fit an empirical relation between the $K_s$ absolute magnitude, metallicity, and radius of M dwarfs. Using this relation, we predict a radius of 0.326 \rsun, in line with our model predictions.
These authors find a $2.7\%$ scatter in M dwarf radii around this relation, and a $1.8\%$ scatter in M dwarf masses.
We consider these values to be extra systematic uncertainties above the model predictions and add them in quadrature to our derived parameters. From this, we find a radius for \thisstar\ of $0.330 \pm 0.010$ \rsun\ and a mass of $0.331 \pm 0.009$ \msun, which are the values we employ through the remainder of this work. We also calculate the density of \thisstar\ to be $9.3 \pm 0.8$ $\rho_{\odot}$, where the uncertainty mostly arises from the aforementioned scatter in M dwarf radii and masses.

\begin{deluxetable*}{lcc}[!ht]
\tablecaption{Stellar information relevant to follow-up efforts. Stellar mass, radius, and density are derived as described in Section~\ref{sec:gaia}. \label{tab:star}}
\tablehead{
\colhead{Parameter} & \colhead{Value} & \colhead{Source} 
}
\startdata
RA &   08:40:06.424218 & \citep{2018Gaia} \\
Dec & +19:05:34.429575 & \citep{2018Gaia} \\
$m_V$ & $16.25\pm0.01$ & \citep{2013Zacharias}\\
$m_J$ & $12.183\pm0.021$ & \citep{2003Cutri} \\
$m_H$ & $11.605\pm0.018$ & \citep{2003Cutri} \\
$m_K$ & $11.370\pm0.023$ & \citep{2003Cutri} \\
$T_{\rm eff}$ (K) & $3385 \pm 70$ &\citep{2018Hirano} \\
$M_\star$ (\msun) & $0.331\pm0.009$ & This work \\
$R_\star$ (\rsun) & $0.330\pm0.010$ & This work \\
$\rho_\star$ ($\rho_{\odot}$) & $9.3\pm0.8$ & This work \\
Distance (pc) & $79.31 \pm 0.45$ & \citep{2018Gaia}
\enddata
\end{deluxetable*}

Recently, \citet{2018Parsons} performed a uniform analysis of low-mass stars, noting that slowly rotating M dwarfs are consistent with models, albeit with a large amount of scatter. 
As there is no obvious rotation period signal in the light curve, nor are any flares observed during the campaigns, the star is consistent with being a slowly rotating M dwarf, so we apply the results of the Dartmouth models without including any corrections to its output.

Using the less precise derived stellar parameters from \citet{2018Rodriguez} and repeating this analysis, we find an increase in the derived mass and radius of less than 1\%, smaller than our quoted uncertainties. In this case, the Gaia DR2 parallax combined with the broadband photometry enable a precise measurement of the stellar mass and radius, overwhelming the spectroscopically-derived temperature and log luminosity, which \texttt{isochrones} accepts as priors. The stellar parameters and their uncertainties used in this work are listed in Table~\ref{tab:star}.

\added{Without a measurement of a rotation period, it is difficult to infer an age for the system. However, the lack of an observable rotation signal or significant flares can provide a lower limit on the age. Stars of similar masses in open clusters like the Hyades and Praesepe have in almost all cases few-day rotation periods and large photometric spot-induced variability which is easily detected in \kt\ data \citep{2019Douglas}. From comparisons to the active M dwarfs in these clusters, this system is likely older than $\sim 1$ Gyr.}

\added{Given the nondetection of the rotation period in any of the 75-day \kt\ campaigns, this system likely has a rotation period longer than $\sim 50$ days, placing this target in the ``slow rotators'' category of \citet{2016Newton}. Those authors suggest the slow rotators have ages broadly in the range $5^{+4}_{-2}$ Gyr, which we take as likely representative of the true age of the K2-146 system.} 

\subsection{Location of transits using Kepler}
\label{sec:kepler}

\subsubsection{Data acquisition and search for periodic signals}
\label{sec:periodsearch}

When first examining the system, our goal is to determine the number of significant transit signals and obtain initial estimates of transit times. For this precursory work, we use long cadence data from campaigns 5, 16, and 18 \added{detrended using the \texttt{EVEREST} pipeline \citep{everest}. A brief comparison shows that the lightcurves generated for this system by the K2SFF pipeline\citep{2014Vanderburg} allow transit times to be determined to roughly the same precision. However, \texttt{EVEREST} can be used to detrend short cadence lightcurves, which we expect to be more useful for the transit parameter fitting process described in Section~\ref{sec:lcf}. Thus, we elect to continue our analysis using only \texttt{EVEREST}}. \explain{Paragraph break added because of the additional text inserted above.}

\begin{figure*}[!tbh]
  \begin{center}
    \includegraphics[width=0.98\textwidth, trim={1cm 0.5cm 1cm 1cm}, clip=true]{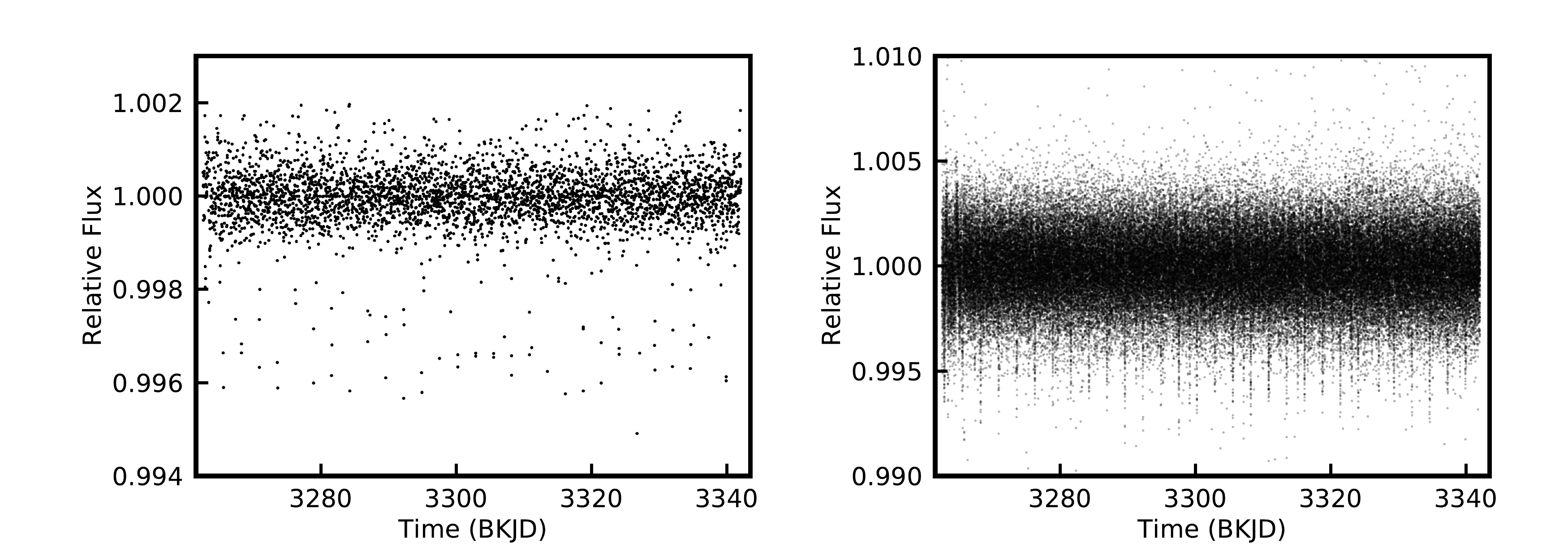}
   \end{center}
  \caption{Long cadence (left) and short cadence (right) lightcurves for K2-146 during C16, detrended and smoothed as described in the text. Some outliers lie outside the range displayed in the right panel, and hence are not shown. These lightcurves, as well as those for C5 and C18, are available online as .fits files.}
  \label{fig:lcurve}
\end{figure*}

We obtain C5 data from the EVEREST database of detrended light curves at the Mikulski Archive for Space Telescopes (MAST) \citep{everest}. The C16 and C18 light curves we download from MAST unprocessed, as those campaigns have not yet been fully processed within the EVEREST pipeline. We use the \texttt{standalone.py} module within the \texttt{EVEREST} Python package to detrend these light curves locally, selecting apertures that encompass the star throughout their respective campaigns and minimize the CDPP (Combined Differential Photometric Precision, a measure of the light curve's noisiness described in \cite{CDPP}). The chosen apertures contained 16 pixels for C16 and 14 pixels for C18, and are not altered during their campaigns. We further detrend and normalize the light curves by dividing each by a running median spanning approximately 2 days. At this point, we note that no data points fall more than 5ppt below unity, while some outliers remain as high as 80 ppt above unity. Thus, we mask out data points \added{above the median by} more than 2ppt, corresponding to 4-5 $\sigma$\deleted{, above the median}. \added{The resulting lightcurve for C16 is shown as an example in the first panel of Fig.~\ref{fig:lcurve}.}

For each campaign, we use \replaced{a}{the} box least squares (BLS) algorithm \added{\citep{refId0}} to search for periodic signals. Initially, we consider 10 transit durations from 1.2 to 4.8 hours and a range of periods from 0.4 d to 30 d. However, when it becomes clear that no significant signals are present at the low or high ends of this period range, we restrict our search to between 1 d and 10 d to save computation time. This process reveals a significant signal near 2.6d and a questionable signal near 4.0d in C5, and significant signals at both stated periods in C16 and C18. While the 4.0d signal in C5 is far from significant, we examine it further because it coincides with significant signals present in C16 and C18. We use these periods to fold the data separately for each campaign, resulting in 6 folded light curves. \added{The Quasi-periodic Automated Transit Search \citep[QATS,]{2013Carter} algorithm is an alternative to BLS that more effectively searches for planets with TTVs, but we choose not to use it because the second signal is noticeable in all 3 campaigns even with just BLS. As will be discussed in Section~\ref{sec:lcresults}, choice of search algorithm may be a less important factor than choice of minimum searched duration in this case.}

\subsubsection{Determination of transit midpoints}
\label{sec:midpointing}

We perform the process that follows for each of these six curves individually. We use the \texttt{batman} Python package \citep{2015PASP..127.1161K}, which analytically calculates transit models using a method detailed in \citep{1538-4357-580-2-L171}, to plot over the top of the folded light curve. To develop an initial estimate for the transit parameters, we adjust five of the model's parameters manually until model and folded lightcurve appear to be in reasonable agreement. The parameters we adjust here are: the ratio between the orbit's semimajor axis and the star's radius ($a/R_\star$), the ratio between the planet's radius and the star's radius ($r_p/R_\star$), the inclination angle of the orbit ($i$), and two limb-darkening parameters (LDPs) for a quadratic limb-darkening model ($u_1$ and $u_2$). At this point, the orbit is taken to be circular, although this constraint is removed later in our analysis. Using the period estimate provided by the BLS search and phase information from the folded curve, we select a window of data points around each transit. For each window, we compute 1000 \texttt{batman} models with midpoint times spaced evenly throughout the window. For each midpoint time, we compare the model and data within the window and calculate the resultant likelihood. We take our prior distribution for each transit's midpoint time to be flat, so our posterior distribution is directly proportional to the calculated likelihoods. We take the midpoint time with maximum likelihood to be our point estimate of a given transit's true midpoint time. We assume the likelihoods to be Gaussian, and thus divide their full widths at half maximum by 2.355 to estimate the uncertainties in the transit times. In cases of multimodal posteriors, we use the widest pair of points at half maximum in order to avoid underestimating transit time uncertainties. We repeat this process---using the new transit times to fold the data, allowing for better guesses of \texttt{batman} parameters, leading to more accurate transit times---and find that after three iterations we cannot noticeably improve the fit simply adjusting the transit parameters manually.

To improve the fit from here, we perform a more careful analysis while switching to short cadence data for C16 and C18 for greater temporal resolution. Because the short cadence data is prohibitively large for detrending in full, we split it into pieces roughly 2.7d long that consist of a similar number of data points as a full campaign of long cadence data. We detrend these pieces individually with \texttt{EVEREST} while masking data around the transit times obtained previously. Because further detrending is heavily impacted by outliers, we mask out points more than about 7 ppt above the highest point or below the lowest point of a running median spanning a campaign with a width of 1.5 hours. 7 ppt is about $4\sigma$ for the short cadence data and large enough that we don't risk eliminating transits. We calculate the best-fit 4th-order polynomial for each piece while masking out the transits, and use these polynomials to detrend and normalize the fluxes. \added{The second panel of Fig.~\ref{fig:lcurve} displays the resulting lightcurve for C16.} We also improve our C5 lightcurve by performing the same piecewise polynomial detrending on the masked post-\texttt{EVEREST} C5 long cadence data. We calculate the residuals between the folded, normalized light curves and the transit models and then repeatedly mask out points with residuals more than 3 standard deviations from zero until no 3-$\sigma$ outliers remain. \explain{Paragraph break added because of the additional text inserted below.}

Next, we use \texttt{scipy} to minimize $\chi^2$ between data and transit model by optimizing twelve parameters: $a/R_\star$ for both planets, $r_p/R_\star$ for both planets, inclination angle for three campaigns each for both planets, and two LDPs. \added{We choose not to use flux contamination as an additional parameter, as the stars in the Gaia DR2 catalog that are brightest and nearest to K2-146 would cause negligible contamination of our apertures. Performing another $\chi^2$ minimization while using the eccentricities and arguments of periastron obtained in Section~\ref{sec:ttvresults} and allowing flux contamination to vary confirms that contamination is insignificant compared to the uncertainty in the transit depths.} Finally, we improve our measurements of the transit midtimes using the updated light curves and transit parameters, following the same procedure described above. \added{The median absolute difference between the new midtime measurements and the previous ones is 1.4 minutes; the short cadence lightcurves, updated transit parameters, and improved midtime measurements lower the reduced $\chi^2$ by 1.3.}

\deleted{The K2SFF pipeline \citep{2014Vanderburg} is an alternative detrending pipeline that also produced long cadence light curves for this system. A brief comparison shows that the K2SFF lightcurves allow transit times to be determined to roughly the same precision as the \texttt{EVEREST} lightcurves. Since we expect the greater temporal resolution of the short cadence \texttt{EVEREST} lightcurves to make them more useful for the transit parameter fitting process described in Section~\ref{sec:lcf}, we choose to continue our analysis with \texttt{EVEREST} lightcurves.}

The transit times we obtain from this analysis are not constrained to physically feasible orbits. Thus, we perform a preliminary TTV analysis similar to the one described in Section~\ref{sec:tta} to obtain a set of physically sensible transit times. For this preliminary analysis, we fit to the photometrically-derived transit times and rate of change of the outer planet's impact parameter. The times obtained in this way are more precise because they align with smooth ephemerides, while the times obtained by individually fitting transits to a transit shape have freedom to align with photometric noise.

\subsection{Lightcurve Fitting} \label{sec:lcf}

We use an affine-invariant Markov chain Monte Carlo algorithm (AIMCMC, \cite{goodman_weare_2010}) via \texttt{emcee} \citep{2013PASP..125..306F} to obtain posteriors on 26 parameters that influence the shapes of the planets' transits. We do this by computing \texttt{batman} models for each planet and campaign and comparing them with lightcurves folded on the dynamically-constrained transit times. The use of an MCMC method allows us to explore 26 degrees of freedom and form quantitative posteriors near the best-fit in a computationally feasible manner.

The stellar fit parameters are $\rho_\star/\rho_{\odot}$ and two quadratic LDPs (3 parameters). Rather than fitting eccentricities $e$ and arguments of periastron $\omega$ themselves, we fit $e\sin{\omega}$, and $e\cos{\omega}$. We fit impact parameter $b/R_\star$, $e\sin{\omega}$, and $e\cos{\omega}$ separately for both planets and all three campaigns (18 parameters). Each planet's radius is fit in the form of $r/R_\star$ (2 parameters). Finally, the relative flux uncertainty $\sigma$ is fit separately in each campaign to ensure we neither overestimate nor underestimate them (3 parameters). \added{We again choose not to include flux contamination as a parameter.} Because the planets' \added{precession-averaged} periods are known precisely, we fit $\rho_\star/\rho_{\odot}$ and calculate both $a/R_\star$ values from it. $b/R_\star$ can be constrained by photometry because of its close relation to transit depth. To generate a transit model, \texttt{batman} requires $i$, $e$, and $\omega$. While we don't fit for those parameters explicitly, we can calculate $e$ and $\omega$ from $e\cos{\omega}$ and $e\sin{\omega}$; we calculate $i$ using the relationship

\begin{equation}
\label{eqn:impact}
\frac{b}{R_\star} = \frac{a\cos{i}}{R_\star} \frac{1-e^2}{1+e\sin{\omega}}.
\end{equation}

\noindent
where $\omega = 90^{\circ}$ when the transit occurs at periastron. We choose to fit $e\sin{\omega}$ and $e\cos{\omega}$ instead of $e$ and $\omega$ because the transit duration is primarily affected by $e\sin{\omega}$. This can be seen from the following expression for the transit duration as a fraction of orbital period:

\begin{equation}
\label{eqn:dur}
\frac{\tau}{T} = \frac{\sqrt{(r_p+R_\star)^2-b^2}}{\pi a}\frac{\sqrt{1-e^2}}{1+e \sin{\omega}}.
\end{equation}

\noindent
(In this paper, transit duration always refers to the time between beginning of ingress and end of egress, since other definitions can become undefined for grazing orbits.) The second eccentricity component, $e\cos{\omega}$, plays only a minor role, and is largely unconstrained photometrically when an artificial timing offset is corrected for (see the Appendix for a description of this effect).

We utilize a Gaussian prior on $\rho_\star/\rho_{\odot}$ based on the results of Section~\ref{sec:gaia}. We find that using priors on the LDPs based on the results of \cite{claret} does not significantly alter the best-fit values for the other parameters, so we use flat priors for limb-darkening as well (constraining to physical situations, e.g. negative limb brightnesses are disallowed). We place flat priors on $b/R_\star$ and restrict them to be less than 1.1 to ensure the planets transit. We restrict $e\sin{\omega}$ and $e\cos{\omega}$ to be between -0.3 and 0.3, as we don't expect larger eccentricities to be dynamically feasible. Further, we impose a log-likelihood penalty for each planet and campaign equal to the logarithm of eccentricity, which effectively converts our priors from flat in $e\sin{\omega}$ and $e\cos{\omega}$ to priors flat in eccentricity and $\omega$. Lastly, we use priors flat in $r/R_\star$ and $\log_{10}{\sigma}.$

For the AIMCMC analysis, we create 400 walkers with starting parameters randomly and independently chosen from uniform distributions spanning the values that seem most likely. These walkers are evolved for 10500 steps of burn-in, allowing them to explore the region of phase space that yields reasonable fits. We verify convergence of the walkers to the posterior distribution for all 26 parameters using the test of \cite{geweke_1992}. We evolve them for a further 10000 steps to collect samples. For each parameter, the autocorrelation function of each walker's value is averaged together to obtain a total autocorrelation function for that parameter. The number of steps required for it to decrease to $1/e$ is taken as the autocorrelation length. The autocorrelation lengths of the 26 parameters are found to be between 650 and 900, giving us an effective minimum of 11 independent samples per walker, for a total of 4400 samples.. We present corner plots of these posteriors in the Appendix (figs.~\ref{fig:corner inner} and \ref{fig:corner outer}), and highlight selected results in Section~\ref{sec:results}.

\subsection{Transit Timing Analysis} \label{sec:tta}

We perform N-body simulations to fit the transit midtime data and impact parameter shifts. These integrations use the GNU Scientific Library's implementation of the Prince-Dormand method (gsl\_odeiv2\_step\_rk8pd) to integrate Newton's equations of motion for 3 bodies. The stellar mass is taken at the best-fit value of $0.331 M_\odot$ and not varied, but for the purpose of computing durations the radius $R_\star$ is taken as a fitting parameter and is constrained by the value of $\rho_\star$ derived from the lightcurve shape. The orbital parameters ($P$, $T_0$, $\sqrt{e} \cos \omega$, $\sqrt{e} \sin \omega$, $i$, and $\Omega$) --- a priori taken as uniformly distributed --- are defined at a time in the center of the short cadence data, namely $BKJD=3370$. Each planets' mass ($m_p$) and radius ($R_p/R_\star$) are also fitting parameters. The planets' orbits are evolved backwards in time to find mid-transit times during C16 and C5 and forwards to find mid-times in C18. We hone in on mid-times by a Newton-Raphson technique described by \cite{2010Fabrycky}. When the sky-projection of the separation vector and the velocity vector are perpendicular, we record \deleted{the} them to yield the impact parameter (in $AU$) and the transit durations; these are averaged over campaign to compare with the lightcurve fits. 

These theory values are compared with the transit timing data via a $\chi^2$ statistic. An additional likelihood value is taken from the correlation matrix of the shape parameters. The full likelihood constraining the solutions is: 

\begin{equation}
    \Delta {\cal L} = \exp(\chi^2/2) + \exp( - \onehalf \vec{x}^t {\bf A}^{-1} \vec{x} )
\end{equation}

Here $\vec{x} = \vec{x}_{\rm model}-\vec{x}_{\rm data}$ is the difference between the parameters predicted by a model to the parameters that best-fit the data; these are given in Table~\ref{tab:achparams}. The covariance matrix ${\bf A}$ we used to constrain the model was computed from 363638 samples of the shape model. Due to the low impact parameter allowed for planet b in campaign 5, with a probability indifferent to different models near $b=0$, half of the values of that parameter were flipped negative, so that its best value is taken to be $0$ with Gaussian profile that better matches the actual samples.

The median values and uncertainties on parameters are determined through Differential-Evolution Markov-Chain Monte Carlo (DEMCMC, \citealt{2006TerBraak}). Forty walkers are initialized at the best-fit that was found by a Levenberg-Marquardt algorithm, with separations in the high-dimensional space by small amounts. At the heart of the algorithm is choosing what step to try for each parameter, to produce the next generation of parameter values. To make that choice for each walker, two other walkers are chosen randomly, and their separation in each parameter differenced. That difference vector is multiplied by a number $\gamma r$, that makes the step more conservative. The value $r$ is chosen randomly each draw. The value $\gamma$ is common for all walkers, but it is a function of generation number. It starts at $2.38/\sqrt{2N} = 0.4$, where $N=17$ is the number of dimensions, but it is multiplied each generation by either $0.9$ or $1.1$ to make the steps more or less conservative, based on whether fewer or more than $23\%$ of the past generation's proposed steps were accepted. In this way, we observed $\gamma$ to settle to, then vary around, $\sim0.1$. With each walker having been assigned a trial parameter set, the model is computed and then the step is either taken if the likelihood either increased or it decreased with the ratio of the new likelihood to the old likelihood that was greater than a draw from a uniformly-distributed random number. For jumps that are not accepted, that walker gives another copy of the current state in the chain.

Given the small differences in initialization of parameters, the first steps are very conservative and most proposed jumps are accepted, leading to the separations of the walkers growing exponentially until they span the region allowed by the data, including its covariances. The values of all parameters are plotted versus generation, with initial ones discarded as a burn-in, and the latter ones validated as fair samples via measuring the autocorrelation. In different parameters, the autocorrelation length was between \replaced{320}{450} and \replaced{900}{1650} steps, and we have completed \replaced{21300}{98000} steps after a burn-in of 2000 steps (for each of the 40 walkers), giving at least \replaced{946}{2376} effectively independent samples of the posterior. \replaced{We expect this procedure yields good sampling of the tails at $\sim 2\sigma$ and some sensitivity to the $\sim 3\sigma$ extremes.}{Using the relationship between effective sample size and tolerance level discussed by \citet{vats}, we expect this procedure to sufficiently sample the parameters' posteriors at tolerance levels between 0.04 and 0.08, corresponding to credibility intervals between 0.96 and 0.92.}

\section{Results}
\label{sec:results}

\subsection{Lightcurve results}
\label{sec:lcresults}

\begin{table}
\caption{Estimates of selected planetary, stellar, and general parameters from AIMCMC fitting of lightcurves.}
\begin{tabular}{ccc} \toprule
 parameter & planet b & planet c \\
 \hline 
$b_{C5} [R_\star]$ & $       0.12^{+0.12}_{-0.08}      $  & $        0.994^{+0.007}_{-0.008}$\\
$b_{C16} [R_\star]$ & $       0.56^{+0.04}_{-0.06}     $  & $        0.900^{+0.011}_{-0.014}$\\
$b_{C18} [R_\star]$ & $       0.53^{+0.06}_{-0.09}     $  & $        0.887^{+0.017}_{-0.022}$\\
$\tau_{C5} [hr]$ & $       1.27 \pm         0.02$  & $        0.51 \pm         0.07$\\
$\tau_{C16} [hr]$ & $       1.366 \pm         0.014$  & $        0.74 \pm         0.02$\\
$\tau_{C18} [hr]$ & $       1.352 \pm         0.016$  & $        0.73 \pm         0.03$\\
$r/R_\star$ & $    0.0582 \pm        0.0006$  & $    0.0606 \pm        0.0011$\\
$a/R_\star$ & $      16.84 \pm       0.4$  & $      22.1 \pm       0.6$\\
$(e\sin{\omega})_{C5}$ & $       -0.02 \pm         0.04$  & $        -0.015 \pm         0.18$\\
$(e\sin{\omega})_{C16}$ & $       -0.23 \pm         0.04$  & $        0.04 \pm         0.06$\\
$(e\sin{\omega})_{C18}$ & $       -0.21 \pm         0.05$  & $        0.08 \pm         0.08$\\

 \hline
$\rho_\star/\rho_{\odot}$ & $    9.1 \pm        0.7$\\
$u_1$  & $0.36 \pm 0.15$\\
$u_2$  & $0.25 \pm 0.23$\\
$u_1+u_2$  & $0.61 \pm 0.10$\\
$\log(\sigma_{C5})$ & $       -3.443 \pm         0.017$\\
$\log(\sigma_{C16})$ & $       -2.776 \pm         0.004$\\
$\log(\sigma_{C18})$ & $       -2.741 \pm     0.005$\\
  \hline\hline
\end{tabular}
\label{tab:achparams}
\end{table}

\begin{figure}[!tbh]
  \begin{center}
    \includegraphics[width=0.44\textwidth, trim={0cm 0cm 1cm 1cm}, clip=true]{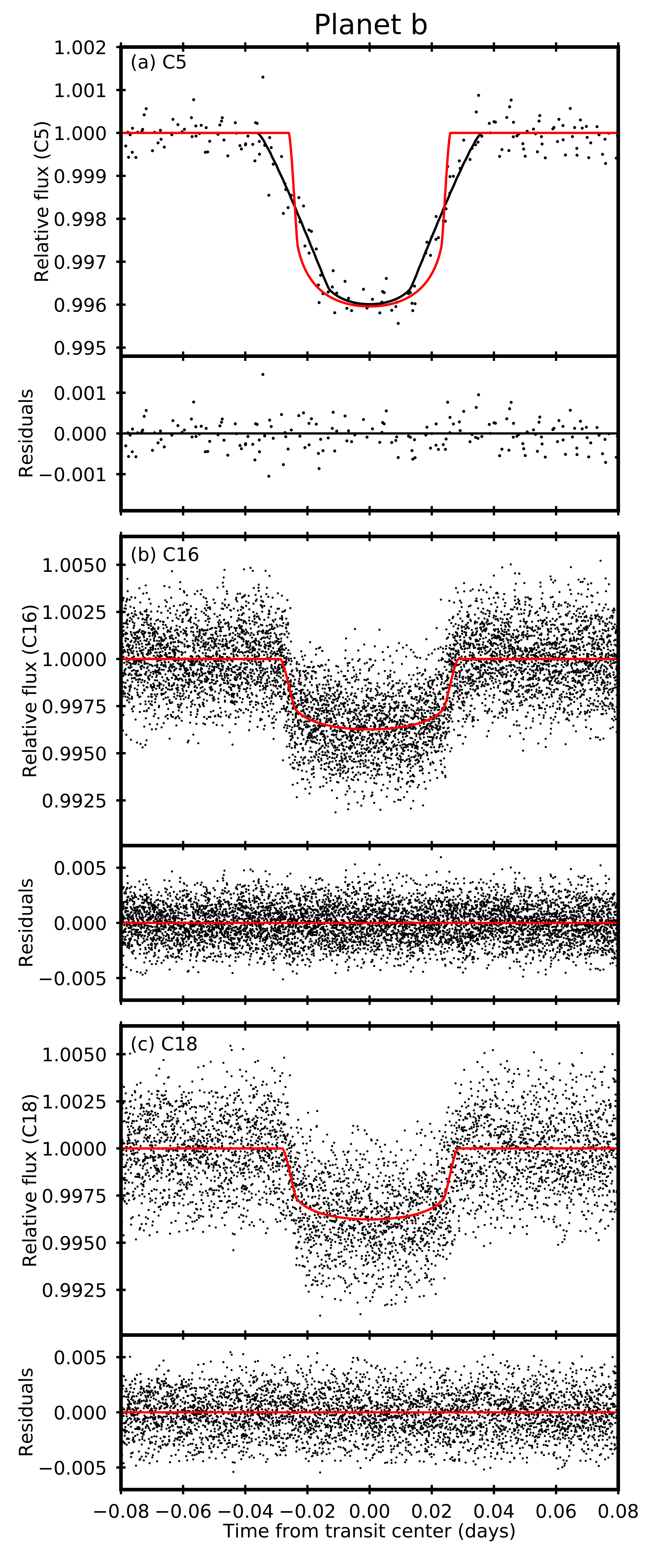}
   \end{center}
  \caption{Lightcurves of planet b, showing data as dots and unbinned \texttt{batman} models as colored lines. (a) shows long cadence data from campaign 5 and also includes the binned model in black, while (b) and (c) show short cadence data from campaigns 16 and 18, respectively. Note that the y-axis scales are different for (a) than for (b) and (c).}
  \label{fig:innerfold}
\end{figure}

\begin{figure}[!tbh]
  \begin{center}
    \includegraphics[width=0.44\textwidth, trim={0cm 0cm 1cm 1cm}, clip=true]{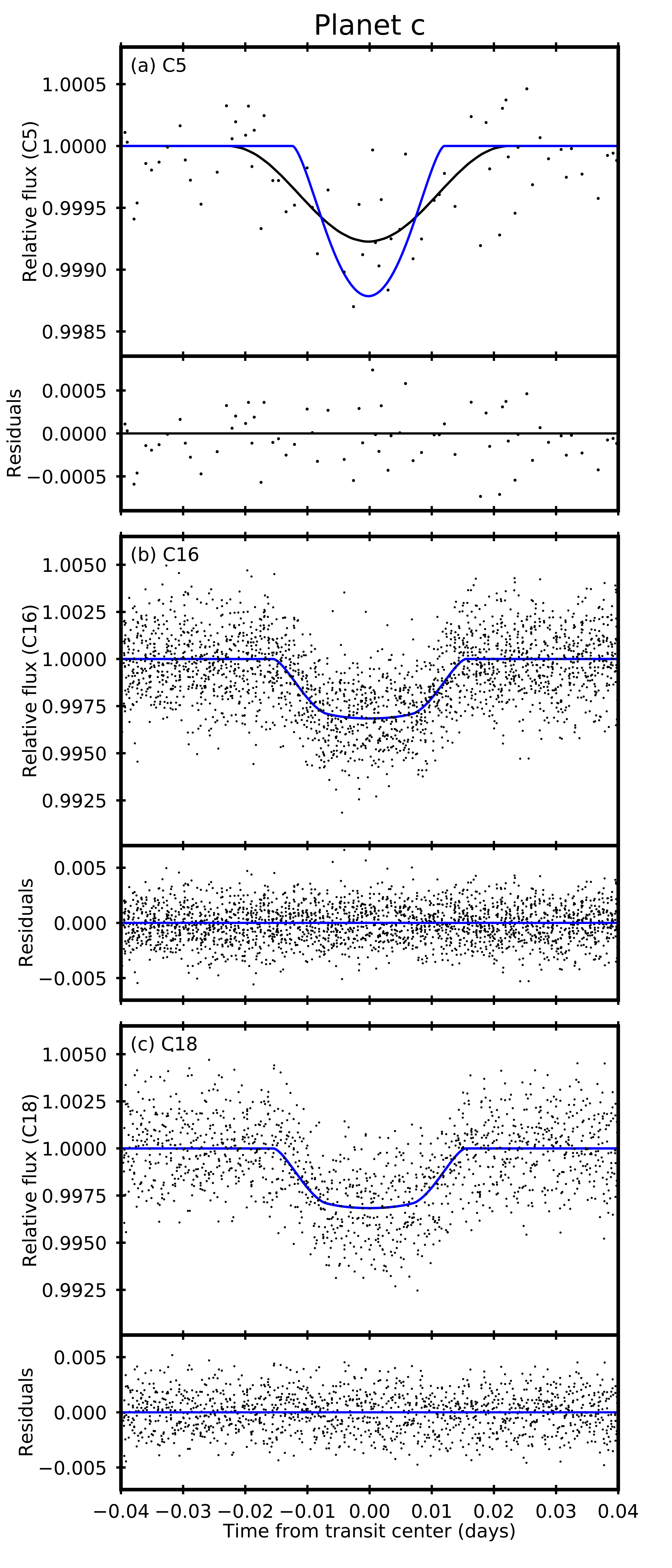}
   \end{center}
  \caption{Analogous plots to those shown in Figure~\ref{fig:innerfold}, except relating to planet c rather than planet b. Compared to Figure~\ref{fig:innerfold}, the x-axis has been rescaled due to planet c's shorter transits.}
  \label{fig:outerfold}
\end{figure}

Light curves folded on the transit midpoints obtained via orbit modeling are plotted alongside best-fit transit models in  Figures~\ref{fig:innerfold} and \ref{fig:outerfold}. The two figures correspond to the two planets, and the subplots to the three campaigns of data. In both cases, the colored lines \replaced{model}{represent} unbinned data, while the black lines \replaced{model}{represent} data binned in 30-minute intervals. Since the duration of a short cadence is much shorter than the duration of a transit, the unbinned models are suitable for comparison to short cadence data. The binned models are plotted only for C5, where they provide a more realistic comparison to the long cadence data than the unbinned models do.

We list our photometry-based estimates of the transit parameters in Table \ref{tab:achparams}. Note that these are not our final parameter estimates; those are updated with the dynamical simulations and instead presented in Section~\ref{sec:tta}. Our estimates of the LDPs are consistent with the values we derive from \cite{claret} of $u_1 = 0.36\pm 0.08$ and $u_2 = 0.35\pm 0.06$. We find that the sum of the LDPs is more well-determined than either of their values individually. This is likely because their sum directly determines the limb darkening at an impact parameter of $R_\star$, and thus is constrained by the depths of planet c's transits. Our fitting procedure leads to $\sigma$ values 15-30\% lower than the median relative flux errors listed in the \texttt{EVEREST}-generated lightcurves. This corresponds to a difference in $\log(\sigma)$ of 0.07-0.14, a highly significant difference.

\begin{figure}[!tbh]
  \begin{center}
    \includegraphics[width=0.44\textwidth, trim={0cm 1cm 1cm 0cm}, clip=true]{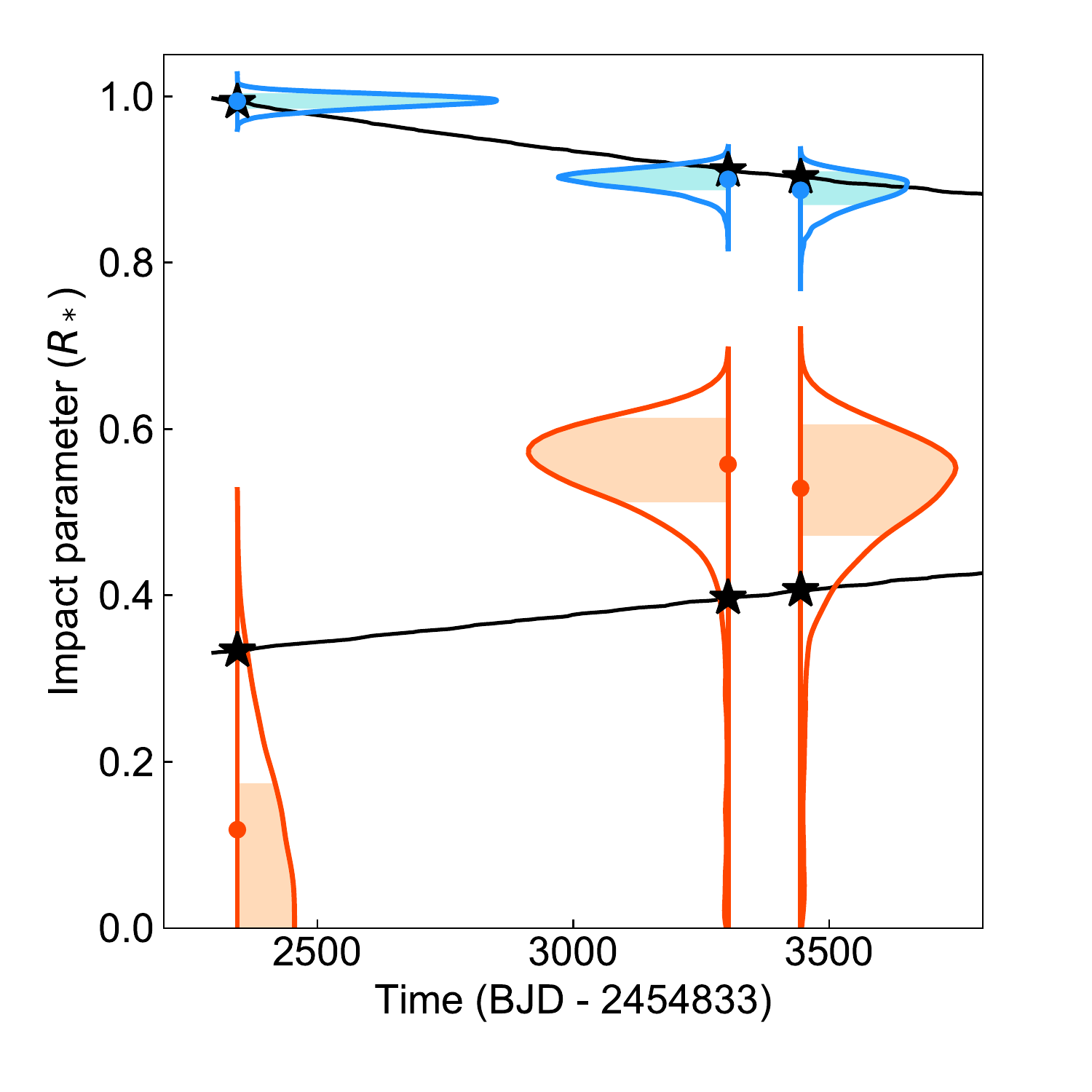}
   \end{center}
  \caption{Half-violin plots of impact parameter during each campaign for both planets. Each half-violin depicts the density of our posterior samples from MCMC fitting of lightcurves --- width is directly proportional to density, with a different scaling used for the two planets for the sake of legibility. The orange plots are for planet b and the blue for planet c. The colored circles are plotted at the mean time within each campaign and the median impact parameter values, while the lightly shaded regions cover 68\% confidence intervals. The black curves show impact parameter as a function of time for the best fit dynamical simulation described in Section~\ref{sec:tta}, and the black stars highlight the impact parameters yielded by that simulation for each planet and campaign.}
  \label{fig:tvb_stack}
\end{figure}

\begin{figure}[!tbh]
  \begin{center}
    \includegraphics[width=0.44\textwidth, trim = {.1cm, .0cm, 1.5cm, 1cm}, clip=True]{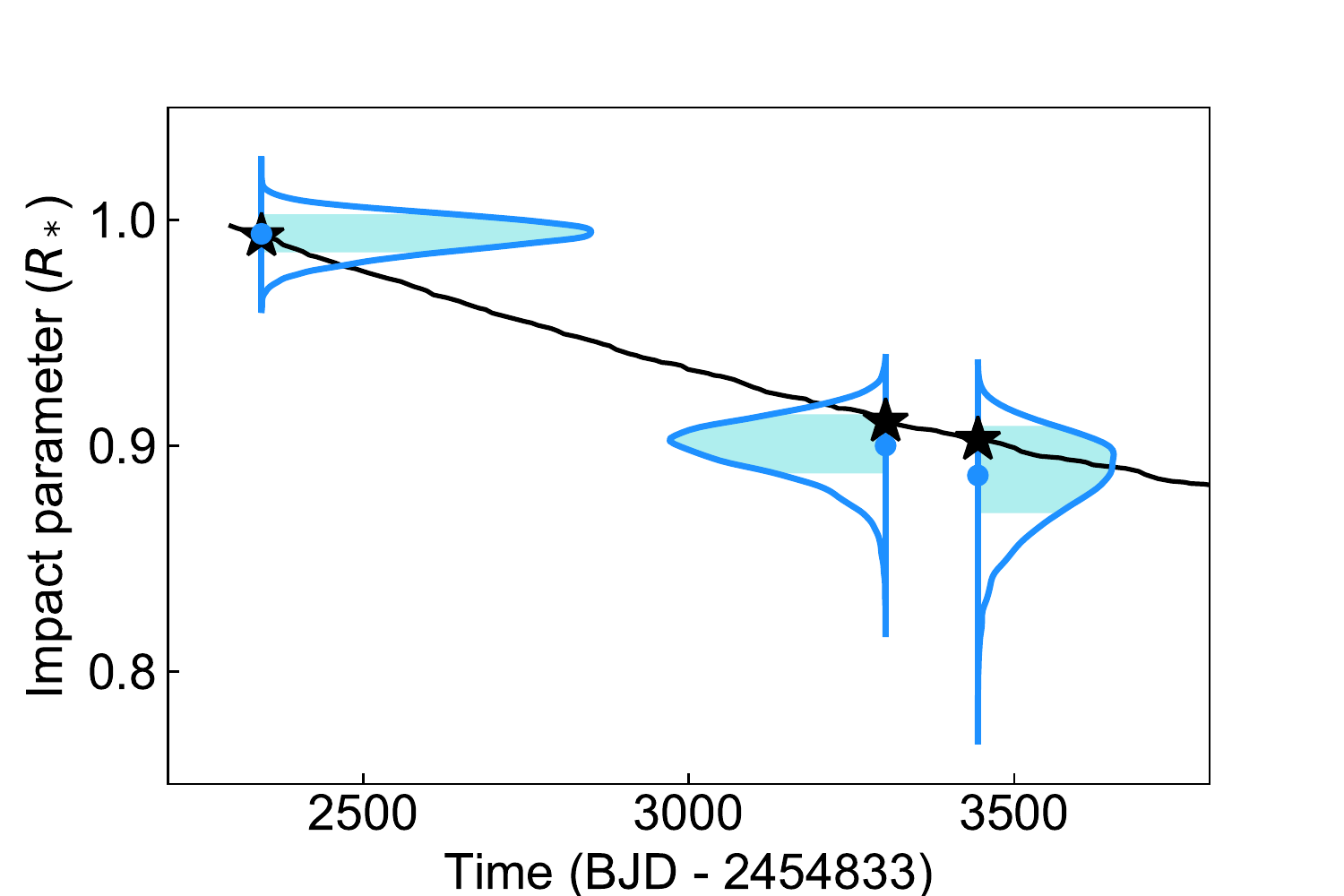}
   \end{center}
  \caption{The same information shown in Figure~\ref{fig:tvb_stack}, zoomed in to make the distributions for planet c easier to see.}
  \label{fig:tvb_czoom}
\end{figure}

Our photometric posteriors for impact parameters are shown in more detail in Figures~\ref{fig:tvb_stack} and \ref{fig:tvb_czoom}. These distributions provide significant evidence that the impact parameter of \deleted{the} both planets have changed over time, and this is reflected in the results of the dynamical simulations of Section~\ref{sec:tta}. The use of these impact parameter distributions as priors in our fitting of the dynamics constrains the system to configurations that yield large ($\sim 0.07-0.1 R_\star$) changes in impact parameters over the 3-year baseline between C5 and C18. Impact parameters near $R_\star$ are much easier to constrain via depth and duration of transit, which is why the impact parameter of planet c is far less uncertain than that of planet b. We note that for planet c to not be grazing during C5, our estimated impact parameter would need to be incorrect by nearly 7$\sigma$.

Planet b's impact parameter is constrained to grow because its transits are deeper during C5 than during C16 and C18. Notably, its transits are briefer during C5 despite its lower impact parameter causing it to cross a longer chord of the star. This constrained $e\sin{\omega}$ to be negative for planet b in C16 and C18 because transit durations are inflated near apastron for eccentric orbits, as can be seen from equation \ref{eqn:dur}.

\begin{figure}[!tbh]
  \begin{center}
    \includegraphics[width=0.44\textwidth, trim={.5cm .5cm 1cm 1cm}, clip=true]{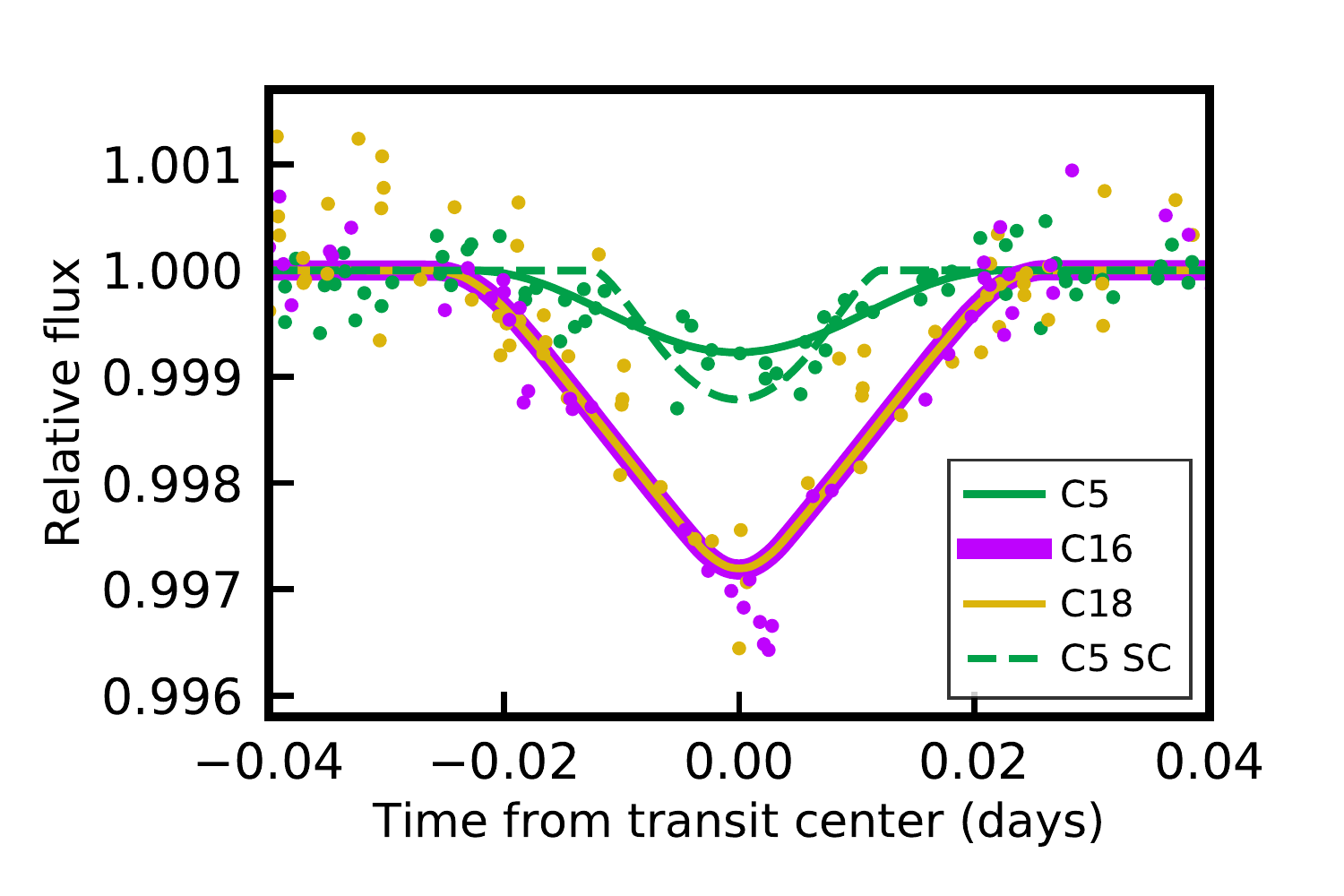}
   \end{center}
  \caption{Long cadence folded lightcurves from each of the three campaigns for planet c. Solid curves show models of the long cadence data\replaced{, with the curve for C16 obscured by the curve for C18.}{. Because the C18 curve falls directly on top of the C16 curve, the C16 curve is enlarged for visibility.} The dashed \replaced{gold}{green} line is a model of what short cadence data would have looked like for C5. The depth change between C5 and C16/C18 long cadence lightcurves is evident. \explain{The colors were changed and the line for C16 was enlarged so that it is still visible despite being behind the line for C18.}}
  \label{fig:binned}
\end{figure}

Figure~\ref{fig:binned} shows observed and modeled long cadence light curves for planet c. The models use the same transit parameters as Figure~\ref{fig:outerfold}, but account for the blurring effect of 0.02-day long cadences. This blurring serves to reduce the depth of the observed signal by approximately 30\% in C5 relative to what would be observed in short cadence data, as can be seen by comparing the solid and dashed \replaced{gold}{green} curves. Even when comparing to C16/18 long cadence data, planet c's transits are 3 times shallower in C5 due to the planet's area partially missing the star during the grazing transit and also to the stronger effect of limb darkening at the edge of the star. These depth-reducing effects contributed significantly to the original nondetection of planet c in the C5 data.

Planet c's detection is also made more difficult by its large-amplitude TTVs and short transit duration. For planets which display a sinusoidally-varying ephemeris with an amplitude which exceeds the transit duration, as is the case for K2-146 b/c, the BLS algorithm can degrade in sensitivity due to the transit-timing variations causing a ``smearing'' of the transit signal \citep{2011Garcia}. In practice this loss of sensitivity is primarily an issue for very shallow transits for which each transit is at or below the noise level, such as planet c during C5. For deeper transits, like planet b, the BLS algorithm can easily identify the linear portions of the ephemeris, with a reduction in the signal-to-noise arising from areas outside of the linear region which only contribute noise.

Finally, planet c's transit duration in C5 of just 30 minutes is comparable to a single long cadence. Most planet searches use grids of transit durations with a minimum of 1.5-2 hours. These searches will therefore have reduced sensitivity to any planets with transit durations much shorter than that minimum search duration. K2's frequent thruster fires caused many one cadence outliers, which likely also made other groups less willing to search for bona fide candidates of such short durations because of the increased rate of false positives. Thus, the nondetection of planet c in all previous searches of C5 can be attributed to its shallow depth, duration shorter than the minimum used in the searches, and TTVs smearing out any remaining signal.

Despite these challenges, a dedicated search would have been able to detect planet c with the appropriate search parameters, most notably using grids of shorter transit durations. As discussed in Section~\ref{sec:periodsearch}, our own BLS search uncovered a tentative signal of planet c in C5 alone using a minimum duration of 1.2 hours --- more than twice the \replaced{planet}{transit}'s actual duration. Kruse et al. (\replaced{2019}{submitted} \explain{The cited Kruse et al. paper is further along the submission process than this Hamann et al. one, and a proper citation can probably be added prior to publication.}) searched the star with an updated version of \replaced{the \cite{2013Carter} Quasi-periodic Automated Transit Search (QATS) algorithm}{QATS} designed specifically to detect small planets with TTVs. While they also found planet b, they missed planet c: again because their minimum search duration of 2 hours was too long. Re-searching the star with their new version of QATS, but with transit durations down to half an hour, recovers planet c in C5 alone, as well as capturing its anti-correlated TTVs to planet b. \added{This suggests that there may be more undiscovered grazing planets in the \kep/\kt\ data that could be detected with searches extending to shorter durations.}

\subsection{TTV results} \label{sec:ttvresults}

\begin{figure}[!tbh]
  \begin{center}
    \includegraphics[width=0.44\textwidth, trim={1cm 3cm 0cm 2cm}, clip=true]{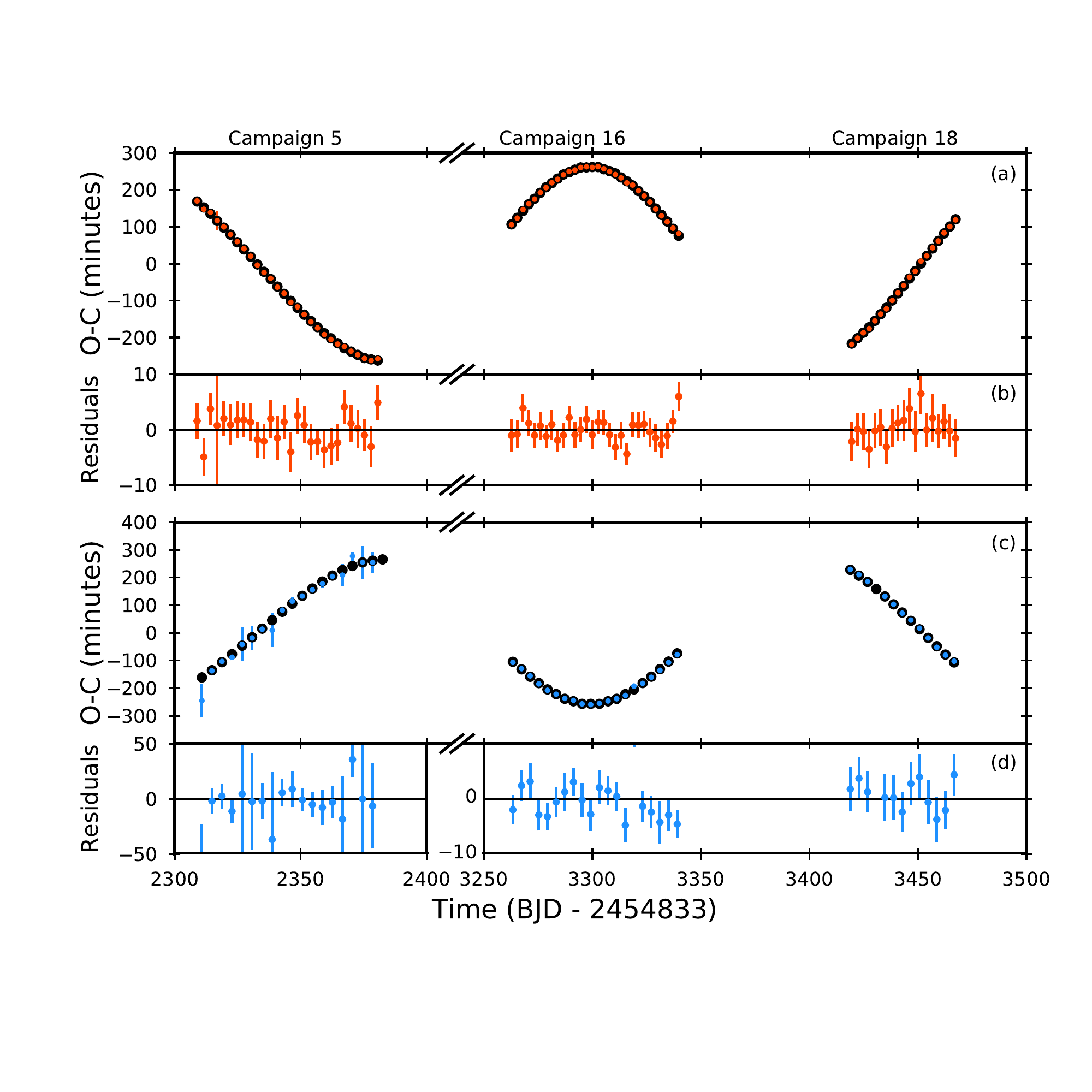}
   \end{center}
  \caption{(a) Modeled (black) and observed (orange) transit times for K2-146\,b
 during the three \kt\ campaigns for which it was observed, relative to a linear
 ephemeris. 
  (b) Residuals between the maximum-likelihood dynamical model and the observed
 transit times. 
  (c) Same as (a), but for K2-146\,c. 
  (d) Same as (b), but for K2-146\,c. For panel (d), note that the scale of the residuals changes
 between Campaign 5 and 16/18 because of the presence of short cadence data and
 the change in impact parameter between these campaigns.}
  \label{fig:ttvdata}
\end{figure}

\begin{figure}[!tbh]
  \begin{center}
    \includegraphics[width=0.44\textwidth, trim={0.5cm 0cm 0cm 0cm}, clip=true]{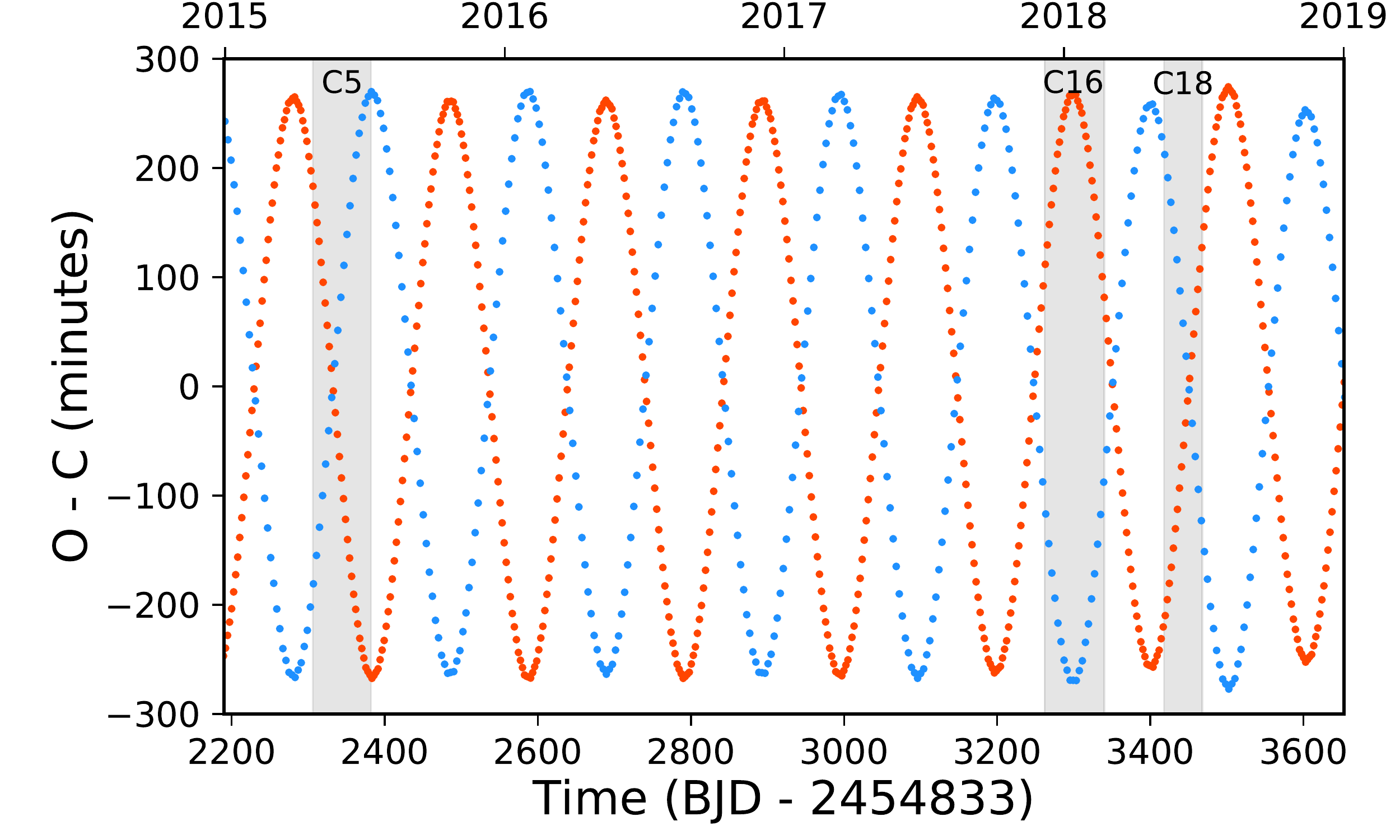}
   \end{center}
  \caption{Maximum likelihood dynamical model for transit times of both planets relative to a linear ephemeris. TTVs for K2-146\,b (c) are shown in orange (blue). The spans in which the star was observed by the \kep\ telescope are highlighted in gray. A clear anticorrelation between the times of transit of the two planets can be seen, \added{causing the ratio between their periods to vary from 1.484 to 1.516 over the TTV cycle. The osculating periods of Table~\ref{tab:params} form a ratio near the top of that range, simply due to the choice of dynamical epoch. The long-term average transit periods (calculated over 1000 years) form a ratio of $\sim1.50025$, which is offset from the exact $3/2$ ratio due to the apsidal precession detailed in section~\ref{sec:dynamics}. } }
  \label{fig:model}
\end{figure}

Here we report results of the transit timing analysis by an N-body code driven by DEMCMC, as described in Section~\ref{sec:tta}, in table~\ref{tab:params} and  Figures~\ref{fig:ttvdata} and \ref{fig:model}. \added{The O-C plots shown in Figures~\ref{fig:ttvdata} and \ref{fig:model} were obtained using periods of 2.65702 and 3.98582 d and $T_o$s of 3371.69669 and 3371.02004 BKJD for planet b and c, respectively.} In the electronic journal article, readers may download chains that are thinned to every \replaced{10}{200} \explain{Using every 10 steps caused the table to be about ten times as large as AAS's MRT prep tool is capable of easily working with. The autocorrelation length is long enough that every 100 steps is still oversampling, so this doesn't reduce the usefulness of the data.} steps, and a brief sample of these is shown in the Appendix in Table~\ref{tab:dansamps}.

\begin{table}
\caption{{\bf Dynamical properties, constrained by transit shapes and times.} The dynamical epoch on the below Jacobian coordinates is \deleted{BJD$-2454833.=$}BKJD=$3370$ \added{(BKJD $=$ BJD$-2454833.$}.  \label{tab:params} }
\centering % centering table
\begin{tabular}{ccc} \toprule
  parameter & planet b & planet c \\
 \hline 
$P$ [days] & $   2.64460 \pm    0.00006$  & $   4.00498 \pm    0.00011$\\
$T_{0} [-2454833]$ & $ 3371.5902 \pm     0.0005$  & $ 3371.1316 \pm     0.0007$\\
$\sqrt e \cos \omega$ & $     0.165 \pm      0.054$  & $    -0.177 \pm      0.034$\\
$\sqrt e \sin \omega$ & $    -0.318 \pm      0.021$  & $     0.200 \pm      0.036$\\
$i$ [deg] & $     88.93 \pm       0.11$  & $     87.54 \pm       0.04$\\
$\Omega$ [deg] & $       0.0 \pm        0.0$  & $       1.9 \pm        0.3$\\
$m_{p} [M_{Jup}]$ & $   0.01816 \pm    0.00024$  & $   0.02358 \pm    0.00031$\\
$R_p/R_\star$ & $    0.0572 \pm     0.0005$  & $    0.0612 \pm     0.0009$\\
 \hline 
 derived:\\
$e$ & $        0.129 \pm         0.019$  & $        0.075 \pm         0.016$\\
$\omega$ [deg] & $ -64 \pm    6$  & $ 130 \pm    5$\\
$m_{p} [M_{E}]$ & $  5.77 \pm 0.18$  & $ 7.49 \pm  0.24$\\
$R_{p} [R_{E}]$ & $ 2.05\pm 0.06$  & $ 2.19 \pm  0.07$\\
\end{tabular}
\end{table}

Correlation plots of each parameter versus the others are given in the Appendix, in Figure~\ref{fig:cornerTTV}.

We fixed $M_\star$ in the dynamical analysis, described in Section~\ref{sec:tta}. In order to fit the shape constraint on $\rho_\star$, we solved for the parameter $R_\star$, as well as $R_p/R_\star$ for each planet. We obtain $\sim 1.0\%$ and $\sim 1.5\%$ precision on $R_p/R_\star$ values of planets b and c. Lightcurves alone can only yield densities, as the equations of motion connect distances with the $1/3$ power of mass. So, after the transit timing fitting, we rejection-sample the joint distribution of $M_\star$ and $R_\star$ obtained in Section~\ref{sec:gaia} based on the updated posterior for $\rho_\star$. Finally, we derive $M_p$ and $R_p$ values from the ratios relative to the star multiplied by the new stellar mass and radius, and propagate the uncertainty from both sources of uncertainty. These planetary masses and radii are given as the last two entries to table~\ref{tab:params}.

\section{Dynamical Evolution}
\label{sec:dynamics}

In this section we explore the dynamical behavior of the system. 

\begin{figure}[!tbh]
  \begin{center}
    \includegraphics[width=0.44\textwidth]{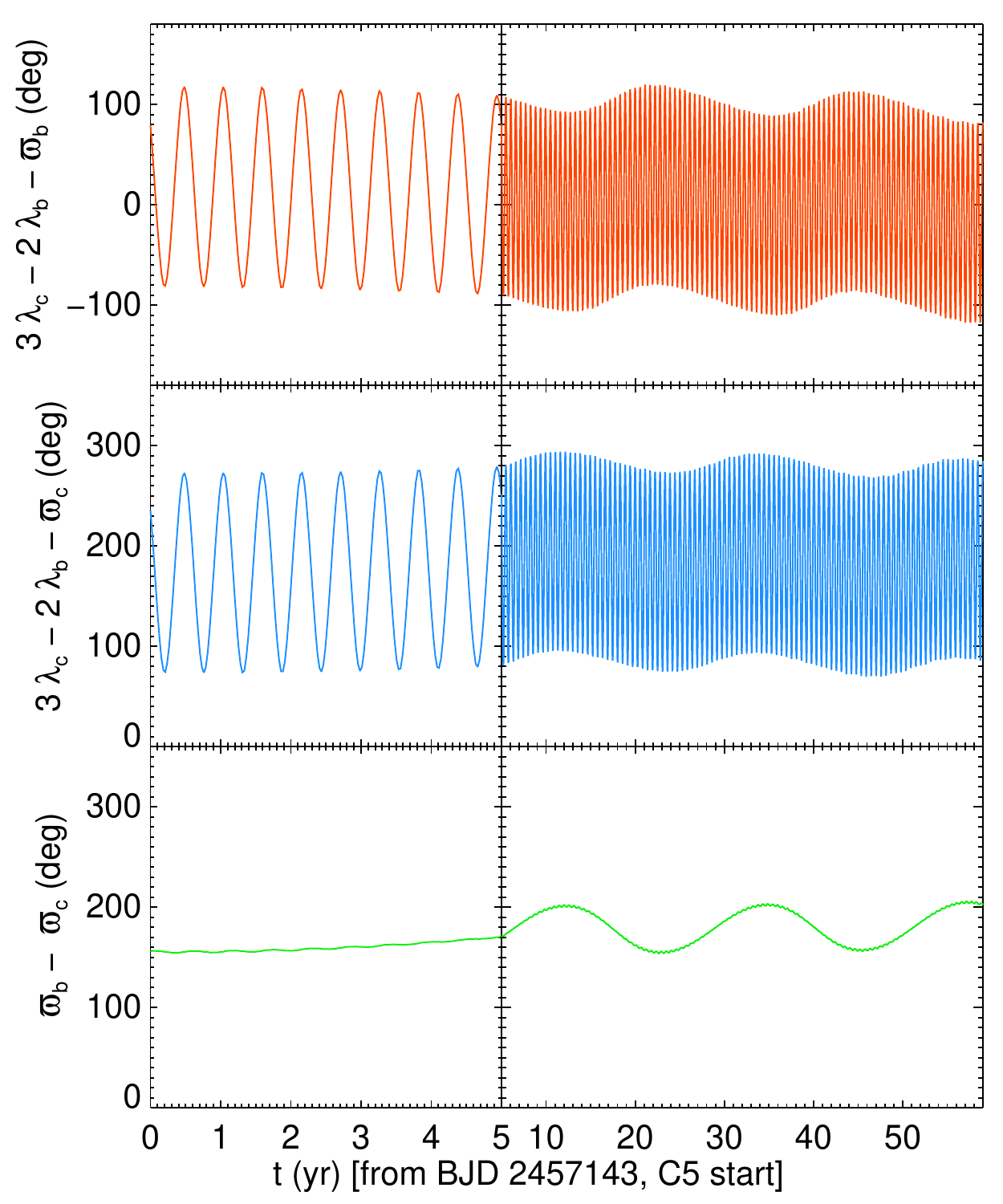}
   \end{center}
  \caption{ {\bf Mean-motion resonance libration}, for one of the draws of the posterior, over 60 years starting from campaign 5. Top two panels are the critical angles for the 3:2 resonance; both librate with large amplitude. The bottom panel is the difference between the apsidal longitudes; libration around $180^\circ$ means the orbits maintain anti-alignment.}
  \label{fig:res}
\end{figure}

The aspect of the dynamics which is linked most closely to the large TTVs is the resonant libration. The planets have resonant arguments composed of their mean longitudes and longitudes of periastron. These oscillate around a mean value, which itself secularly changes (see  Figure~\ref{fig:res}). The long axes of the two orbits librate around an antiparallel configuration. 

The dynamical results require that the two planets are librating in the 3:2 resonance. \deleted{This inference will be important for understanding the dynamical history of the system.} \added{We measured this libration in a subsample of 146 models drawn from the DEMCMC posterior, finding the full-amplitude of libration of $\phi_1=3 \lambda_c - 2 \lambda_b - \varpi_b$ is $200.0^\circ\pm0.8^\circ$, and of libration of $\phi_2=3 \lambda_c - 2 \lambda_b - \varpi_c$ is $199.5^\circ\pm0.5^\circ$. These are the ranges obtained over a single libration cycle.  On decade-long timescales, the libration center moves as in figure~\ref{fig:res}, and the full range visited by the resonance angles widen to $261^\circ\pm22^\circ$ and $246^\circ\pm16^\circ$ for $\phi_1$ and $\phi_2$, respectively.  Meanwhile, $\Delta \varpi$ oscillates about $180^\circ$ with amplitude $85^\circ\pm32^\circ$. Theories for the formation and evolution of the resonance may be constrained based on these results.}

\begin{figure}[!tbh]
  \begin{center}
    \includegraphics[width=0.44\textwidth, trim={.6cm .4cm 1.3cm 1.6cm}, clip=true]{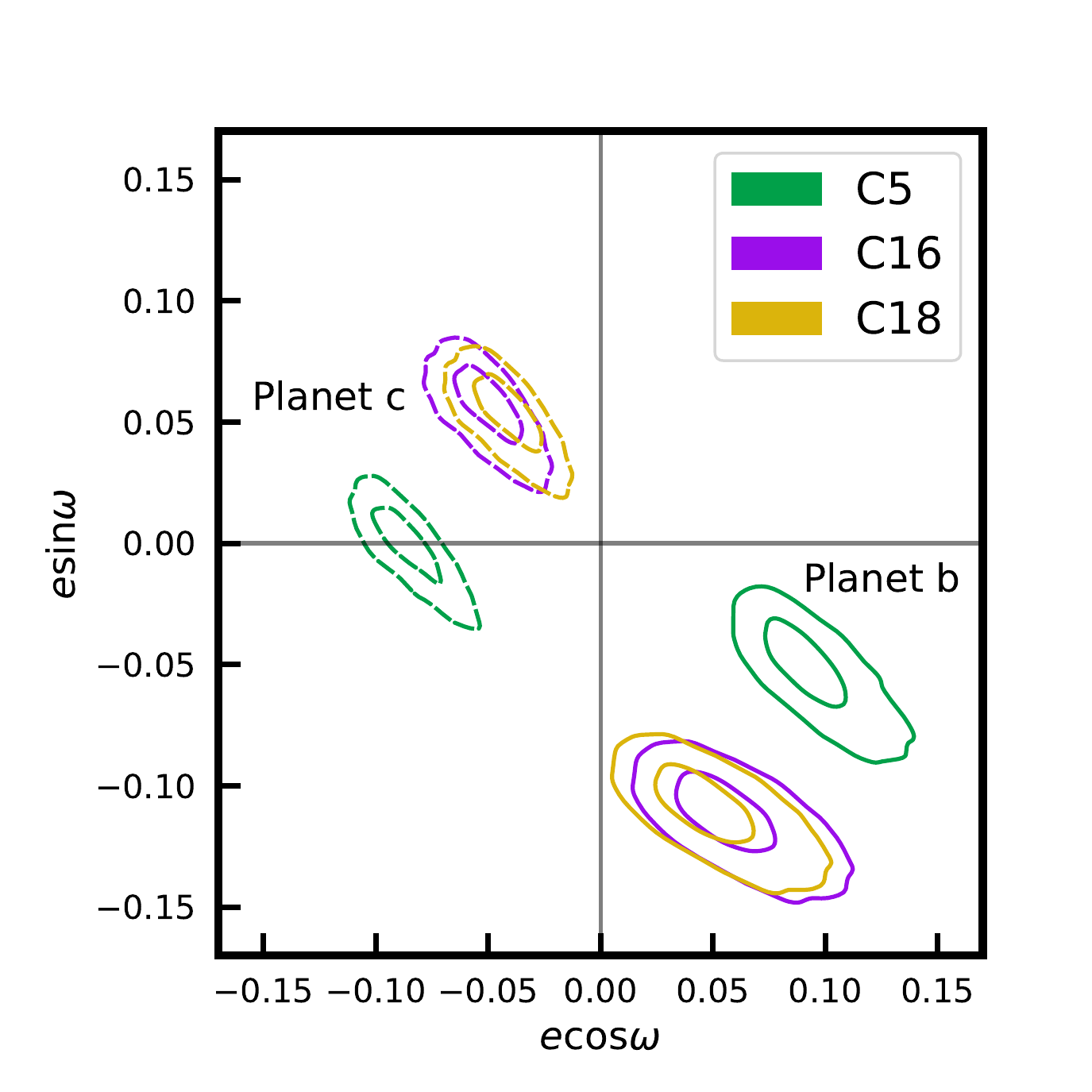}
   \end{center}
  \caption{1-$\sigma$ and 2-$\sigma$ contours for the eccentricity vectors (defined as where the probability density falls to $e^{-1/2}$ and $e^{-2}$, respectively, times the peak density). \explain{The colors were changed so as to remain consistent with Fig.~\ref{fig:outerfold}. The trim settings were changed so the axis labels wouldn't get partially cut off.}}
  \label{fig:ecosesin}
\end{figure}

Figure~\ref{fig:ecosesin} shows the eccentricity components as a function of campaign for each planet. The increase in transit duration of the outer planet is partially due to out-of-plane torque changing the inclination of the planet to the sky plane, and partially due to in-plane precession causing the periastron to swing towards the observer. This latter effect shortens the distance between the planet and the star by about 5\%, and the sky-projected projected component of this separation therefore decreases by 5\%. Thus, the impact parameter \deleted{goes} \added{would have gone} from 1.0 to 0.95 by that effect alone, which dramatically increases the depth of the transit.

\begin{figure}[!tbh]
  \begin{center}
    \includegraphics[width=0.44\textwidth]{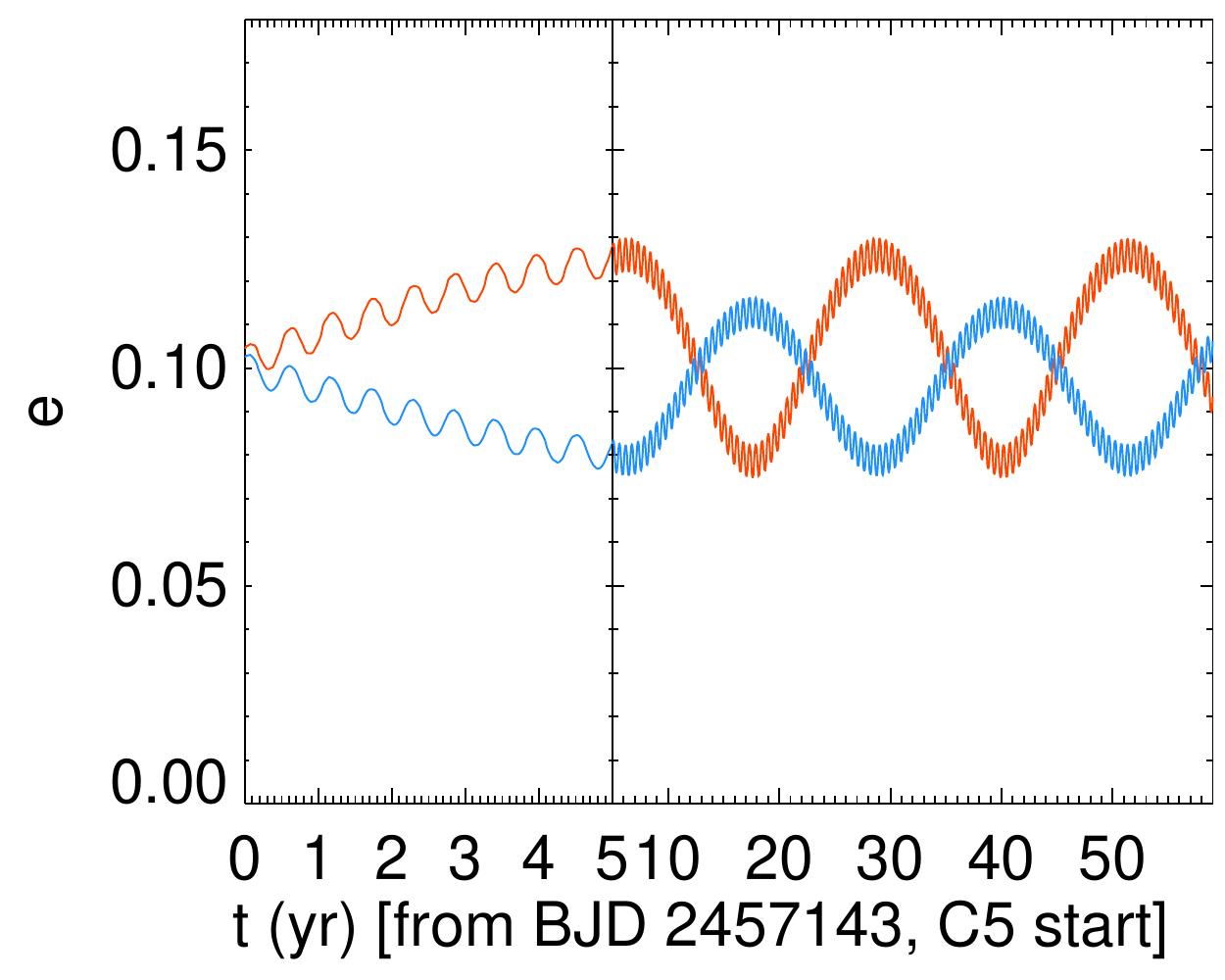}
   \end{center}
  \caption{ {\bf Eccentricity oscillations} of the two planets, on a fast timescale due to the resonant perturbation, and on a longer timescale due to secular perturbation.  }
  \label{fig:e}
\end{figure}

\begin{figure}[!tbh]
  \begin{center}
    \includegraphics[width=0.44\textwidth]{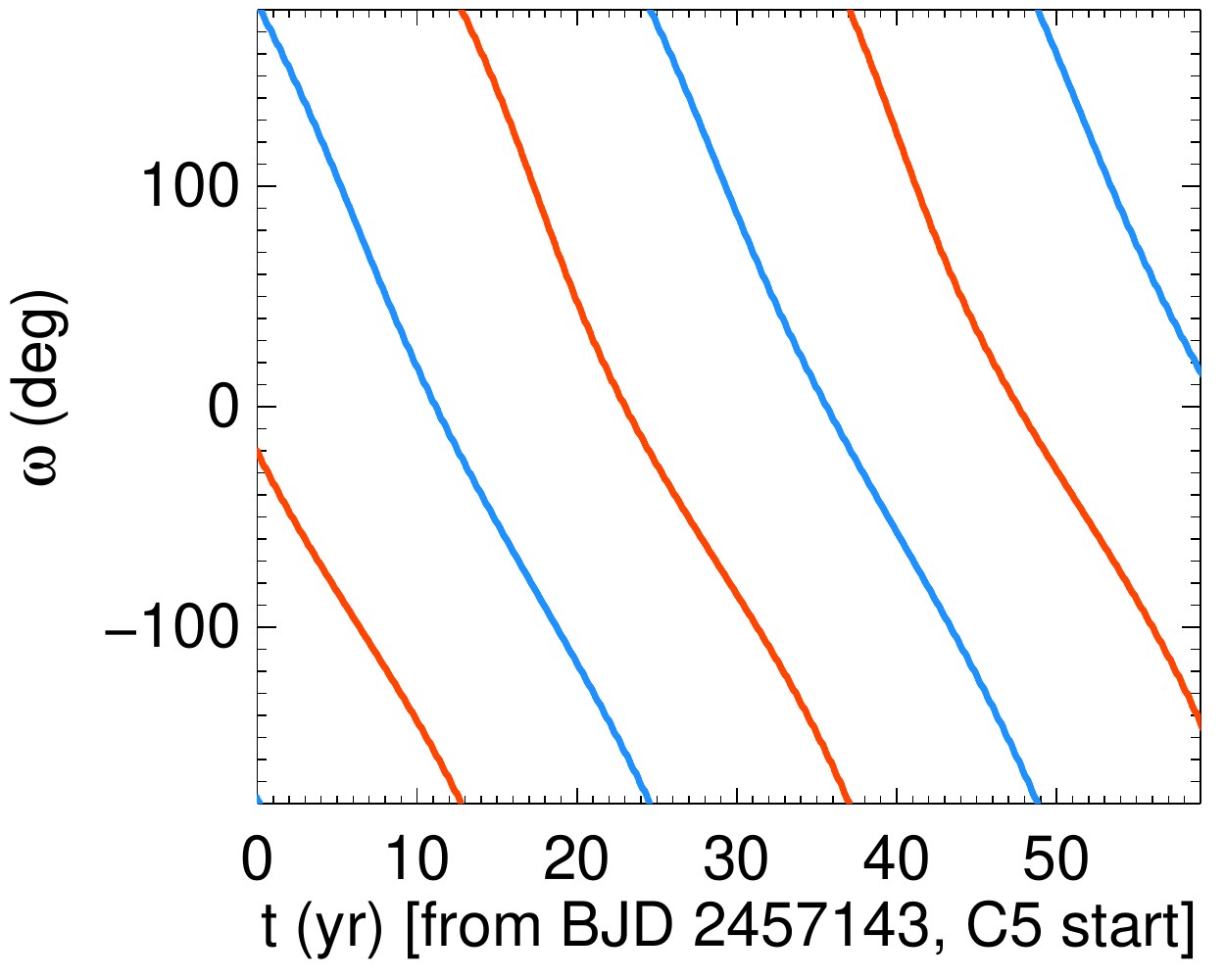}
   \end{center}
  \caption{ {\bf Periastron precession}, one of the two causes of depth variations for planet c, and the cause of the duration variation in planet b.  }
  \label{fig:omega}
\end{figure}

\replaced{Over a timespan longer than the data, the periastra of both planets make complete circuits around the star (Figure~\ref{fig:omega}). }{The eccentricity and the arguments of periapsis of each planet are shown in figures~\ref{fig:e} and \ref{fig:omega}, respectively. Over a timespan longer than the data, the eccentricities variations cause angular momenta to secularly swap between the planets, and the periastra of both planets make complete circuits around the star.}

\begin{figure}[!tbh]
  \begin{center}
    \includegraphics[width=0.44\textwidth]{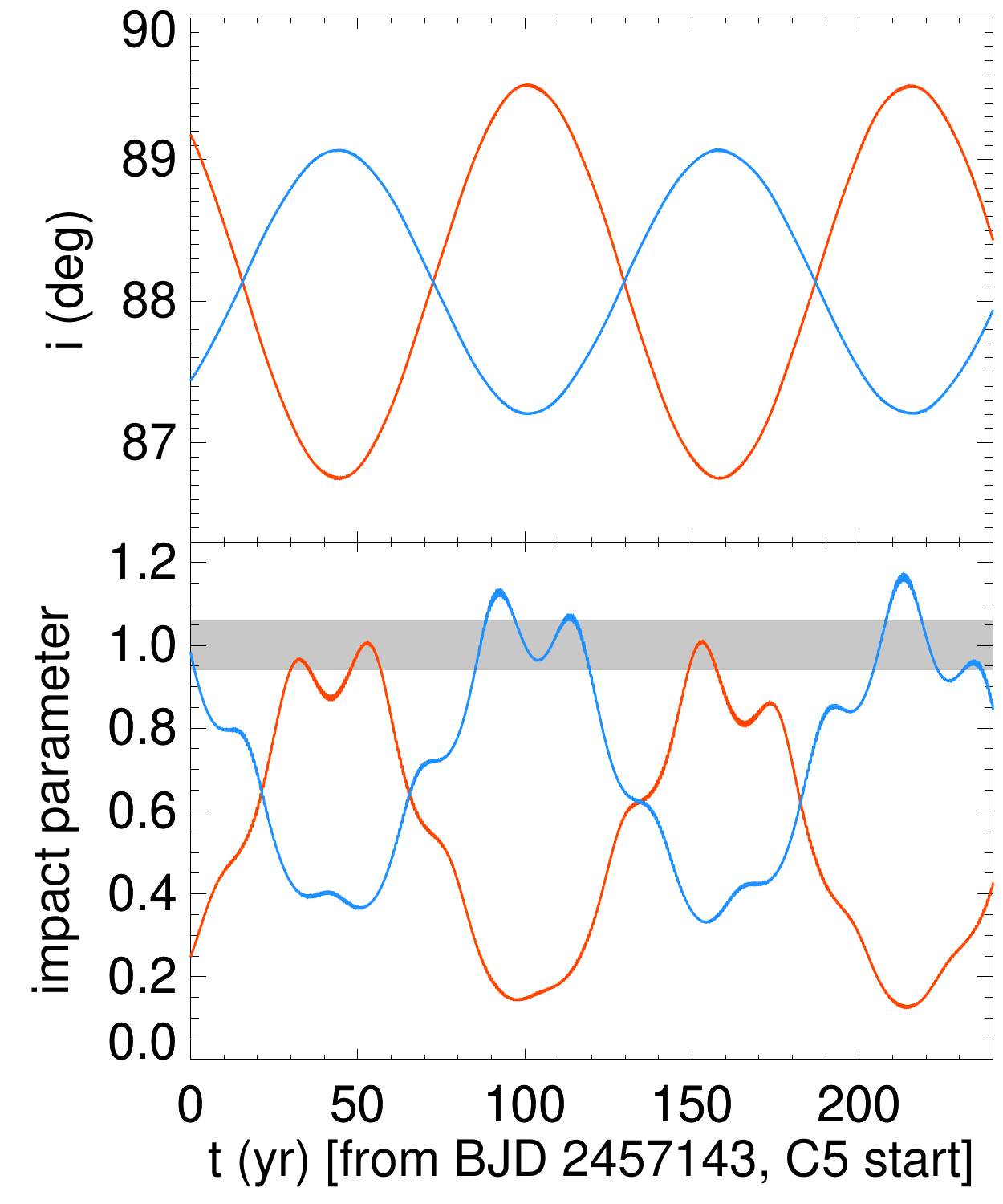}
   \end{center}
  \caption{ {\bf Sky-plane inclination (top) and impact parameter (bottom)} for the two planets, showing a repeating pattern of precession. The bottom plot has short-timescale variations due to in-plane orbital precession of the eccentric orbits, bringing the planets closer to the star (and to lower impact parameter) when the periastron aligns with the line of sight. The gray zone on the bottom plot indicates a grazing transit for $R_p/R_\star=0.06$, approximately the value of these planets. On a short timescale, both effects combine to cause planet c's transit depth and duration to increase.}
  \label{fig:inc}
\end{figure}

The mutual inclination of $2.40^\circ\pm0.25^\circ$ causes a nodal precession with a period of $\sim 106$~years (shown for one model in Figure~\ref{fig:inc}). This rate is much slower than the resonant precession rate (200 days), the 20-year $\Delta \varpi$ and eccentricity oscillation, and the 22-year periastron precession rate. Such a hierarchy of timescales has been noted for other resonant systems as well \citep{2010Correia}.

Figure~\ref{fig:inc} also shows impact parameter variations for the next 300 years, according to one model. It appears that planet c is more likely to be grazing (in the gray region) or missing the star (above the gray region) than planet b. Running 146 of the models described in Section~\ref{sec:tta} forwards 1000 years provided further evidence for this \added{statement}. Across the 146 models, planet c spent an average of  $16\pm2\%$ of the time grazing and $12\pm3\%$ of the time missing; meanwhile, planet b had an average grazing fraction of $10\pm3\%$ and missing fraction of $<5\%$ (at 95\% confidence). Additionally, these results suggest that \deleted{the} it is not rare in general for both of these planets to be observable at once.

We also confirmed that our solutions to the system are long-term stable. Some known systems have poorly characterized libration, or other sensitivity, such that not all solutions to the data are viable models. That is, since we are observing this planetary system that is likely Gyr old, we expect that it is in a stable state, and running it forward a few Myrs would show a regular pattern of motion. We selected 20 draws from the posterior and ran them forward in time with the Mercury integrator. They all survived for 10 Myr. On the one hand, this task suggests that all the models that are allowed by the data are viable; on the other hand, we cannot use a stability criterion to further constrain the system.

\section{Discussion}
\label{sec:discussion}

\begin{figure}[!tbh]
  \begin{center}
    \includegraphics[width=0.44\textwidth]{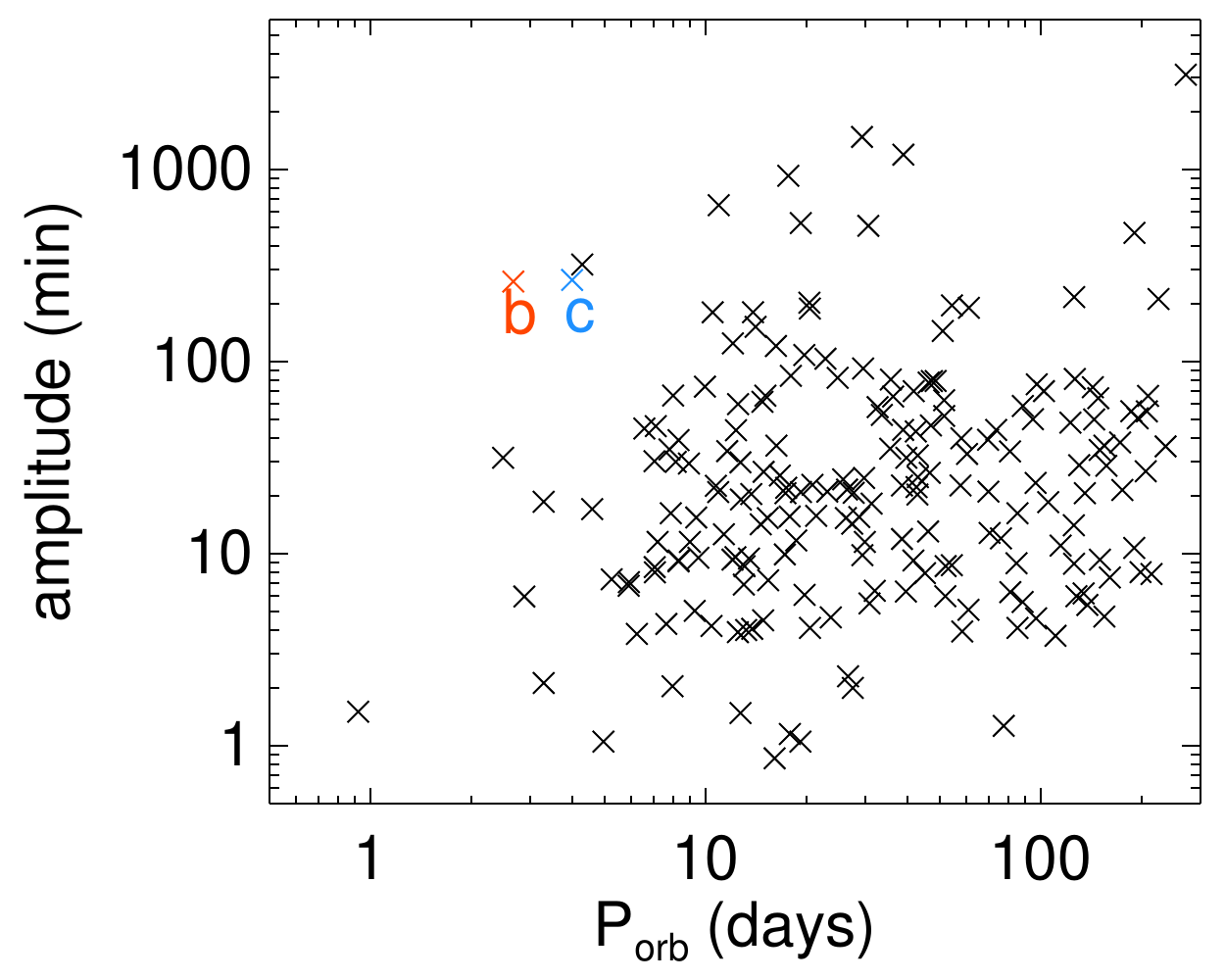}
   \end{center}
  \caption{ {\bf Transit timing amplitudes versus average planetary period.} The positions of planets b and c of the \thisstar\ system are marked. }
  \label{fig:ttvamp}
\end{figure}

The striking transit timing and depth variations of the \thisstar\ system can be compared to what was discovered in the \emph{Kepler} prime mission. No duration changes as dramatic were found for planets orbiting single stars --- only for circumbinary planets, whose precession and moving hosts caused impact parameters to vary from transit to transit. Regarding transit timing, \cite{2016Holczer} have measured the transit times of many of the planet candidates, computing transit timing amplitude of the dominant sinusoidal component in each planet. We compare the distributions of that amplitude and average planetary period to the \thisstar\ planets in  Figure~\ref{fig:ttvamp}.

We see that the planets lie in an unusual spot in this space: being in resonance with large amplitude libration allows a large TTV amplitude to be visible, while having such a small orbital period allows us to see many cycles of the variation. This combination has led to the $\sim1\%$ precision on $M_p/M_\star$ values. This privileged location is shared only with KOI-984.01, which is the only known transiting planet in its system. Thus it is very difficult to solve the KOI-984 system uniquely, and it is impossible to constrain the mass of KOI-984.01 to the level we have constrained the planets of \thisstar.

\subsection{Structure of the K2-146 planets}

\begin{figure}[!tbh]
  \begin{center}
    \includegraphics[width=0.46\textwidth, trim={0.0cm 0cm 0cm 0cm}, clip=true]{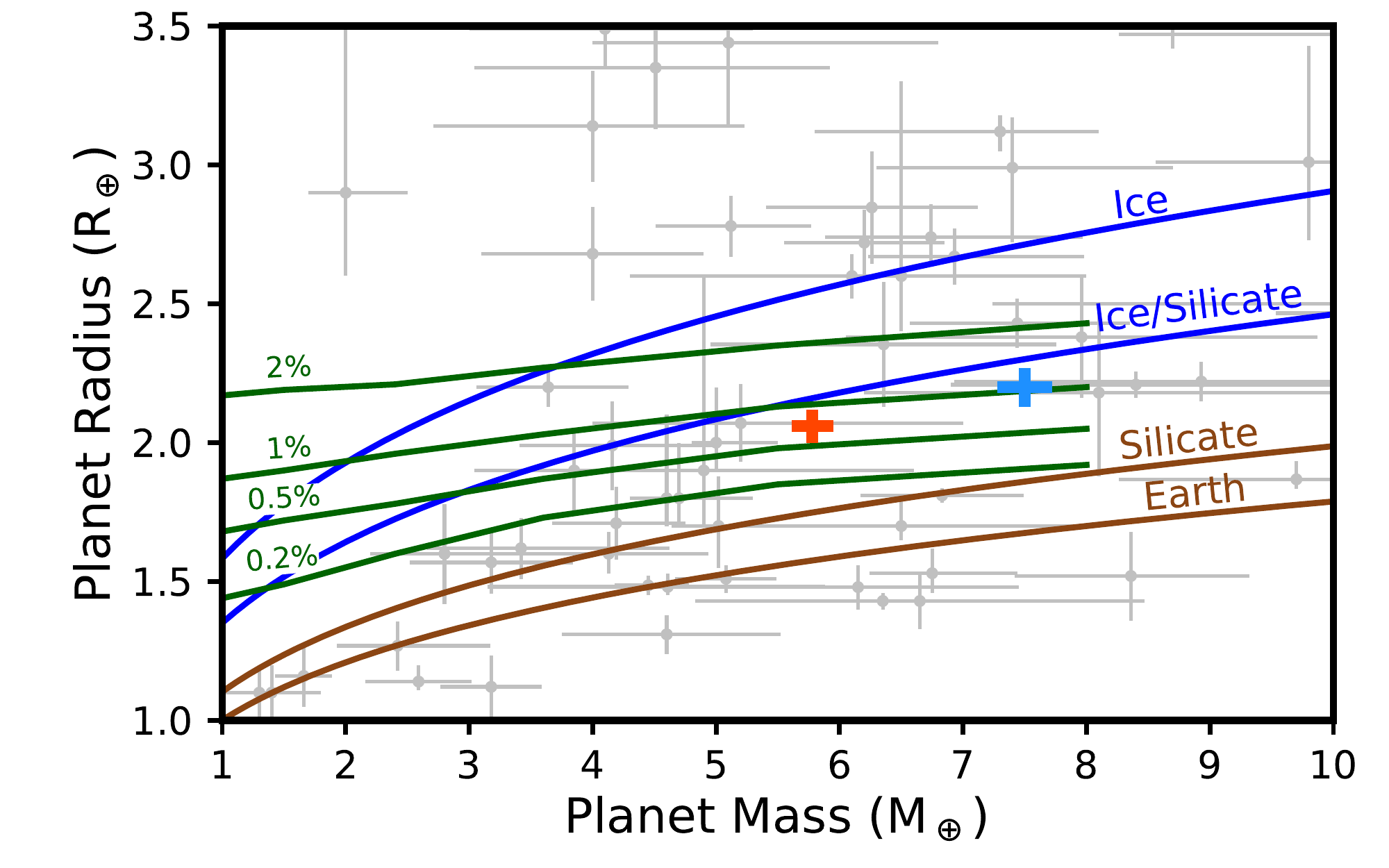}
   \end{center}
  \caption{Mass and radius of \thisstar\,b (orange) and \thisstar\,c (blue). Theoretical models of masses and radii are shown in blue and brown for planets that are (top to bottom) 100\% ice, 50\% each ice and silicates, 100\% silicates, and an Earth-like composition of 67\% silicates and 33\% iron. The curves are taken from \citet{2007Fortney}. In green are models for rocky planets with extended hydrogen/helium atmospheres. From top to bottom, these are 0.2\%, 0.5\%, 1\%, and 2\% gas fraction by mass. Gray points in the background are all planets with both masses and radii measured to be nonzero at $3\sigma$ significance. Both planets are consistent with having equal amounts of ice and silicates; alternatively, both could also be rocky cores with a $\sim1$\% by mass extended gaseous envelope.}
  \label{fig:mr}
\end{figure}

We turn to comparing the mass and radius constraints to other exoplanets and to theoretical models, depicted in Figure~\ref{fig:mr}. The \thisstar  planets' densities of $0.67\pm0.04$ and $0.71\pm0.05 \rho_\oplus$ (or $3.69\pm0.21$ and $3.92\pm0.27 g/cm^3$), respectively, are consistent with being either water worlds or having substantial atmospheres \citep[see also Figure~\ref{fig:mr}]{2007Fortney, 2014Lopez}. At first glance, this might be surprising given their short orbital periods.
\citet{2013Owen} proposed the existence of an ``evaporation valley," suggesting that planets with radii of $\sim 2$ \rearth\ should be rare relative to planets with radius of 1.5 \rearth, which have had their atmospheres photoevaporated away, or 2.5 \rearth, which have strong enough gravity to retain 
their atmospheres.
Such a valley was confirmed by \citet{2017Fulton}; \citet{2018VanEylen} expanded on this result, showing 
that the radius gap is a function of orbital period, in line with the predictions of \citet{2013Owen}.

\begin{figure*}[!tbh]
  \centering
  \begin{minipage}[b]{0.48\textwidth}
    \includegraphics[width=\textwidth]{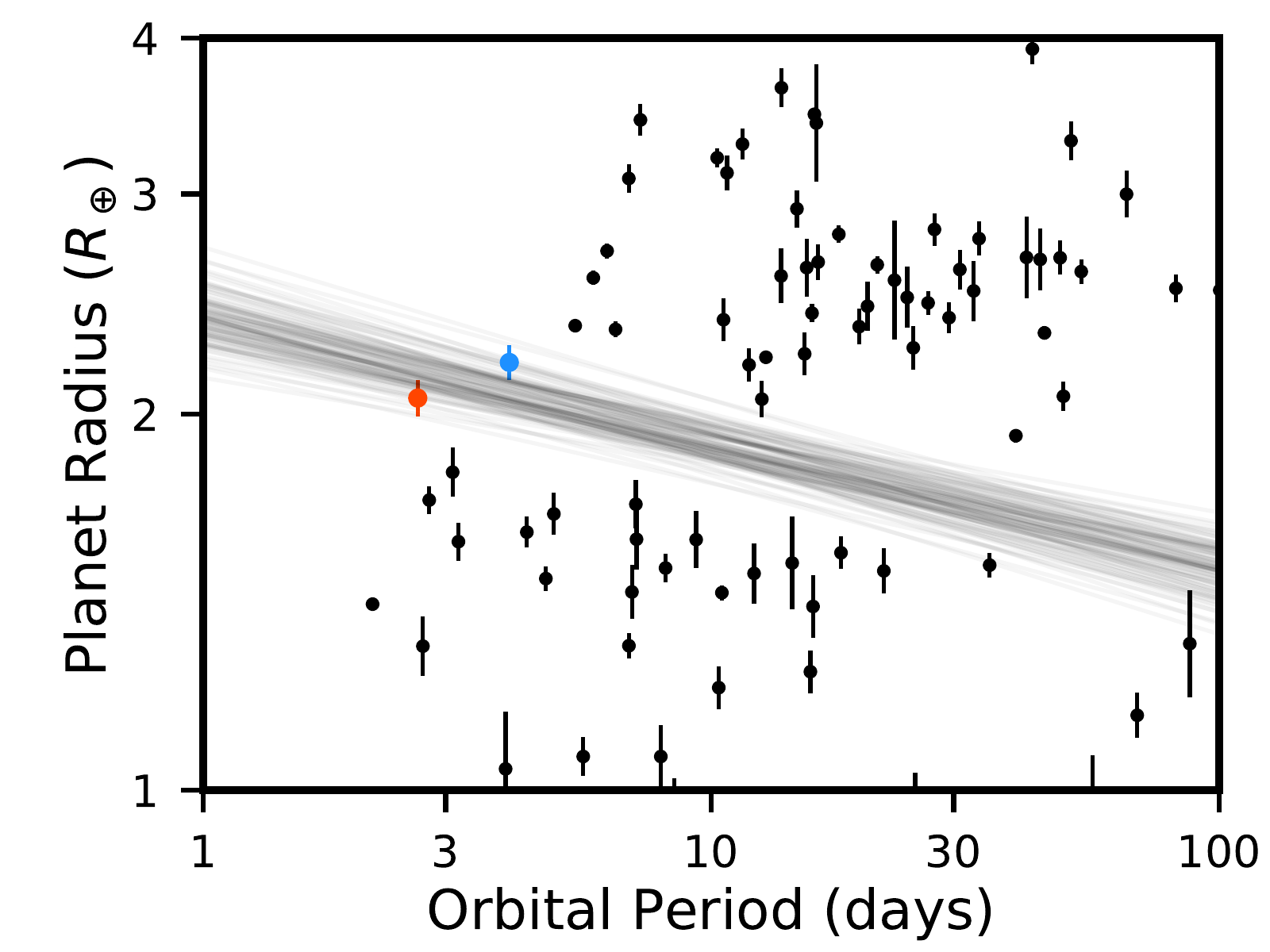}
  \end{minipage}
  \hfill
  \begin{minipage}[b]{0.48\textwidth}
    \includegraphics[width=\textwidth]{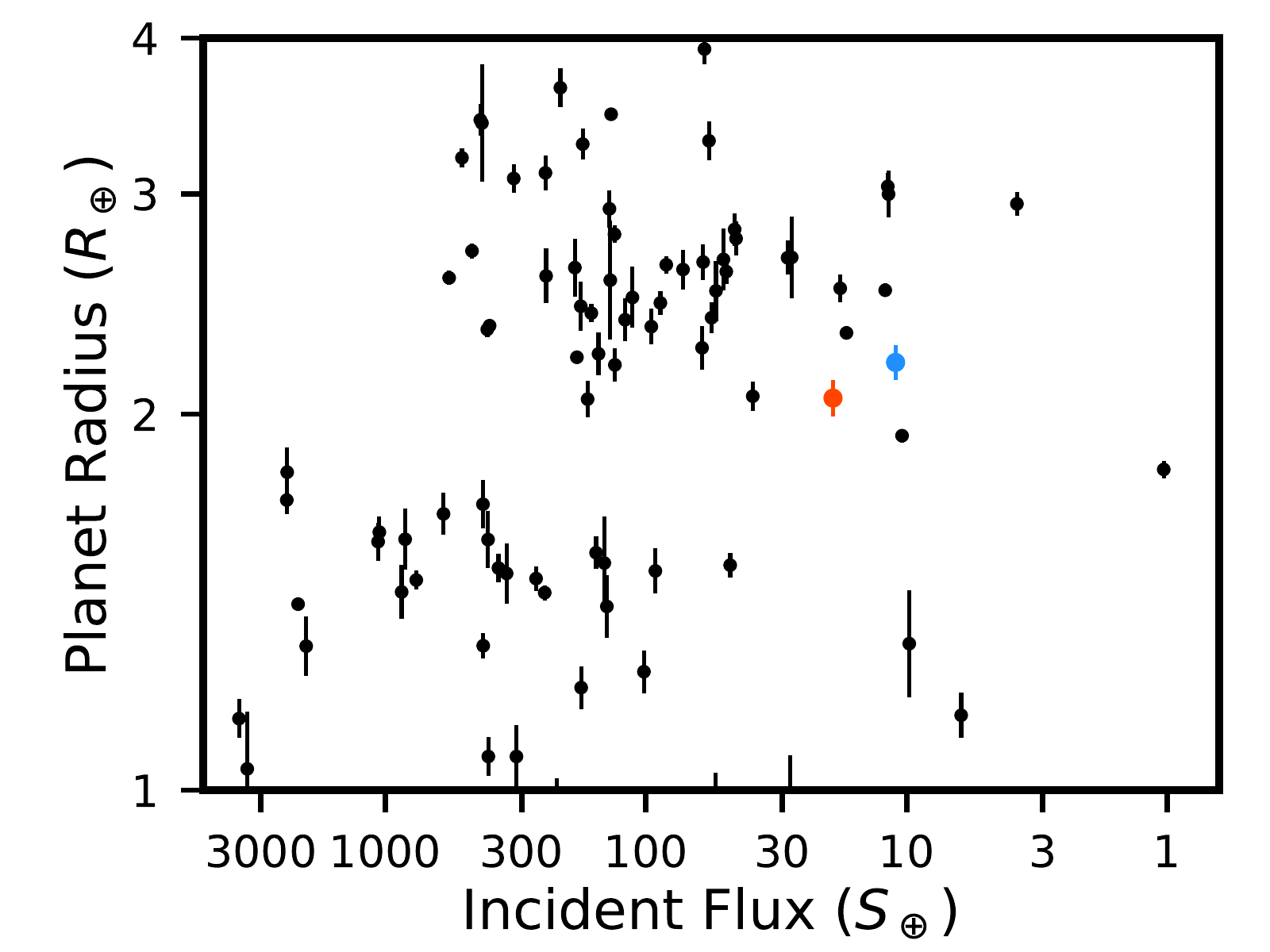}
  \end{minipage}
  \caption{Observed planet radius as a function of (left) orbital period and (right) incident flux, with data taken from \citet{2018VanEylen}. The lines on the left panel represent those authors' estimate of the linear model that describes the 
  ``evaporation valley'' proposed by \citet{2013Owen}. At first glance, the two planets orbiting \thisstar\ appear to fall on
  opposite sides of the photoevaporation gap. However, when presented with respect to
  incident flux, both planets are comfortably above the gap, suggesting they could have significant gas envelopes. The position
  of \thisstar-b on both plots suggests that only orbital period and planet radius alone are incomplete tracers of the photoevaporation gap: the properties of the host star must be considered as well. Our host star is an M-dwarf, in contrast to the Sun-like hosts of all the black points here, which are derived from an asteroseismic sample.} \label{fig:peg}
\end{figure*}

Figure~\ref{fig:peg} shows the orbital period and radius of the \thisstar\ planets relative to the 
evaporation valley as inferred by \citet{2018VanEylen}. From those two parameters alone, it would
appear that \thisstar\,b is below the gap and therefore should have lost its primordial atmosphere
due to photoevaporation. However, these analyses used samples of FGK stars observed spectroscopically and asteroseismic targets, which are mostly more massive than the Sun. 
These planets, orbiting a mid-M dwarf, likely faced a very different UV environment over their first 
100 Myr \citep[e.g.][]{2013Stelzer, 2015Ansdell}. There is a large scatter in the details of UV 
emission from M dwarf to M dwarf \citep{2014Shkolnik}, making it challenging to interpret the early UV environment for these planets. As a simple proxy, we can consider the bolometric incident flux received
by the planet at the present time. If we reframe the evaporation valley in that context, as shown in Fig. \ref{fig:peg}, we see the planets are both comfortably above the observed gap, suggesting both planets 
could retain their atmospheres to the present day. The existence of \thisstar\,b in its present form strongly suggests that the specific stellar environment in which a planet resides should be considered 
when trying to interpret the history of its atmosphere and its susceptibility to photoevaporation.

\subsection{Future Observations}

\thisstar\ is faint ($V=16.2$) in the optical \citep{2012Zacharias}, limiting
the possibility of RV follow-up with many
facilities. 
However, the star is significantly brighter at redder wavelengths \citep[$i=14.2, J=12.2$;][]{2003Cutri, 2012Zacharias}, meaning current and 
planned RV facilities optimized to observe M dwarfs at redder wavelengths could
be used to obtain additional observations of this system.
RVs could also detect the presence of non-transiting planets in this system.
This was the case for K2-18, where follow-up RV observations identified an inner non-transiting companion to a transiting planet \citep{2015Montet, 2017Cloutier}.
Additionally, while giant planets only orbit a few percent of mid-M dwarfs \citep{2014Montet}, the presence of a massive, outer planet could provide clues about
the formation and dynamical evolution of these transiting planets \citep[e.g.][]{2016Otor, 2017Gratia}. 

While the star is faint, the observed transits of the inner planet are 0.4\% in depth, making these transits visible from the ground with modest telescopes. RVs and additional transit photometry both provide future opportunities on reasonable timescales to determine which plausible solution accurately describes the \thisstar\ system. The errors on planet properties are already driven by the uncertainties on stellar characterization. The main motivation to continue measuring the planetary periods is to refine the ephemerides of when future transits will occur. This would be important if a campaign is envisioned for probing the thin atmospheres on these planets. Predicted transit times and uncertainties until September 2022 are available for digital download, and a preview is shown in Tables~\ref{tab:futureb} and \ref{tab:futurec} in the Appendix; ephemerides can be improved most efficiently by observing future transits with the highest timing uncertainties. We expect that both planets would have JWST S/N below the cutoff recommended by \cite{2018Kempton} for atmospheric characterization. However, because their masses are known more precisely than is usually possible with RV measurements, they may be valuable candidates for atmospheric study nonetheless.

\subsection{The \kt\ Legacy Field as a \textit{TESS} Testbed}

\thisstar\ demonstrates the power of the \kt\ mission, and space-based transit missions in general, for detailed characterization of planetary systems through repeated observations of fields. Without any ground-based follow up observations, we have
succeeded in measuring the mass of both planets to a precision of 3\% despite the \kep\ spacecraft observing 
this system for less than 20\% of the observational baseline. 
In this case, the Campaign 5 observations were fortuitously timed so that the presence of TTVs could be inferred from that data set alone, but longer-baseline TTVs in other systems will only be detectable by combining data from Campaign 5, 16, and 18 together. Moreover, continued observations of previously observed fields enable the detection of new, previously undiscovered planets. This is both true in the case of \thisstar\,c, where the planet's transit depth increased between campaigns, and in the likely more typical case where additional observations raise the S/N of a phase-folded transit signal above the criterion for significance. This system, and the \kt\ legacy field in general, thus serves as a dress rehearsal for the possibilities that a \textit{TESS} extended mission will enable, when transits are observed in sectors separated by long data gaps. In future work, we will investigate other systems in this field, including a search for previously undetected planets and TTVs.

\acknowledgements

We thank Leslie Rogers for discussing the nature of gaseous envelopes on small planets and Rodrigo Luger for help with the EVEREST pipeline. \added{We also thank the referee for providing helpful comments which improved the manuscript.}

A.C.H. was supported by a K2GO5 grant from NASA, number 80NSSC18K0364.

Work by B.T.M. was performed under contract with the Jet Propulsion
Laboratory (JPL) funded by NASA through the Sagan Fellowship Program executed
by the NASA Exoplanet Science Institute.

\added{EA acknowledges support from NSF grant AST-1615315 and NASA grants NNA13AA93A and 80NSSC18K0829.}

This paper includes data collected by the \kt\ mission.
Funding for the \kt\ mission is provided by the
NASA Science Mission directorate. We are grateful to
the entire \kep\ and \kt\ team, past and present.

This work has made use of data from the European Space Agency (ESA) mission
{\it Gaia} (\url{https://www.cosmos.esa.int/gaia}), processed by the {\it Gaia}
Data Processing and Analysis Consortium (DPAC,
\url{https://www.cosmos.esa.int/web/gaia/dpac/consortium}). Funding for the DPAC
has been provided by national institutions, in particular the institutions
participating in the {\it Gaia} Multilateral Agreement.

Some/all of the data presented in this paper were obtained from the Mikulski Archive for Space Telescopes (MAST). STScI is operated by the Association of Universities for Research in Astronomy, Inc., under NASA contract NAS5-26555.

\software{%
    numpy \citep{numpy},
    matplotlib \citep{matplotlib},
    lightkurve \citep{lightkurve},
    isochrones \citep{2015Morton},
    astropy \citep{2018AJ....156..123A},
    scipy \citep{scipy},
    kadenza \citep{kadenza},
    everest \citep{2018Luger},
    emcee \citep{2013PASP..125..306F},
    corner \citep{corner},
    batman \citep{2015PASP..127.1161K}
    }

\facility{Kepler, Gaia, Exoplanet Archive}

%\bibliography{papers}

\appendix

\begin{deluxetable}{ccccccccccc}[!ht]

\tablecaption{Planetary, stellar, and noise parameters for best-fitting lightcurve model. \label{tab:achbestfit}}
\tablehead{
\colhead{pl.} & \colhead{$r_p/R_\star$} & \colhead{$b_{C5}/R_\star$} & \colhead{$(e \sin \omega)_{C5}$} & \colhead{$(e \cos \omega)_{C5}$} & \colhead{$b_{C16}/R_\star$} & \colhead{$(e \sin \omega)_{C16}$} & \colhead{$(e \cos \omega)_{C16}$} & \colhead{$b_{C18}/R_\star$} & \colhead{$(e \sin \omega)_{C18}$} & \colhead{$(e \cos \omega)_{C18}$}
}
\startdata
b & 
0.0584872 & 0.0829449 & -0.0063295 & -0.0013517 & 0.5971109 & -0.2887000 & 0.0980958 & 0.5717864 & -0.2867856 & -0.2544630\\
c & 
0.0609491 & 1.0010763 & -0.2172455 & 0.1146076 & 0.9038561 & 0.0052956 & -0.0004055 & 0.9023093 & 0.0134860 & 0.0159588\\
\hline\\
&$\rho_\star/\rho_\odot$ & $u_1$ & $u_2$ & $\log(\sigma_{C5})$ & $\log(\sigma_{C18})$ & $\log(\sigma_{C18})$\\
\hline &
9.8060629& 0.3381495& 0.2524213& -3.4276080& -2.7734142& -2.7472132
\enddata
\end{deluxetable}

\begin{deluxetable*}{cccccccc}[!ht]
\tablecaption{Best-fitting dynamical model, Jacobian coordinates at dynamical epoch BJD 2458203.000 (BKJD $+3370$). Assumed $M_\star=0.331 M_\odot$. The fitting parameter minimized is $(2 \ln \Delta {\cal L}) =163.3337$. Planet masses given in units of $M_{\rm Jup}=9.545\times10^{-4} M_\odot$. \label{tab:bestfit}}
\tablehead{
\colhead{planet} & \colhead{$P$ (days)} & \colhead{$T_0$ (BKJD)} & \colhead{$\sqrt{e} \cos \omega$} & \colhead{$\sqrt{e} \sin \omega$} & \colhead{$i$ (deg)} & \colhead{$\Omega$ (deg)} & \colhead{$M_p$ ($M_{\rm Jup}$)}
}
\startdata
b & 2.64461662195& 3371.59032337250& 0.14179457798 & -0.31149532326 & 88.94655839504  & 0 (definition)  &0.01825083297\\
c & 4.00491983123 & 3371.13139767470 &  -0.19543722063 & 0.21749371377 & 87.51200896614 & 2.01729810288 & 0.02371931622
\enddata
\end{deluxetable*}

\begin{deluxetable*}{ccccccccccccccccc}[!ht]
%\tabletypesize{\scriptsize}
\tablecaption{A preview of samples described in Section~\ref{sec:tta}. The full table is available as a machine-readable table. \label{tab:dansamps}}
\tablehead{\colhead{$\rho_\star (\rho_\odot)$}&
\colhead{$P_b$ (days)}&
\colhead{$T_{ob}$ (BKJD)}&
\colhead{$\sqrt{e_b}\cos{\omega_b}$}&
\colhead{$\sqrt{e_b}\sin{\omega_b}$}&
\colhead{$i_b$ (deg)}&
\colhead{$\Omega_b$ (deg)}&
\colhead{$M_b (M_{Jup})$}&
\colhead{$R_b (R_\star)$}}
\startdata
10.2549796& 
2.64459256& 
3371.59078& 
0.13014153& 
-0.3336921& 
88.7788353& 
0.0& 
0.01821541& 
0.0581626
\\
8.63854893& 
2.64469711& 
3371.59074& 
0.14716385& 
-0.3012654& 
88.9552587& 
0.0& 
0.01833648& 
0.0576053
\\
9.50392865& 
2.64457815& 
3371.59015& 
0.16020293& 
-0.3429077& 
88.8031510& 
0.0& 
0.01813466& 
0.0575424\\
... &
... &
... &
... &
... &
... &
... &
... &
...\\
 \\
 \hline
 \hline
 & 
$P_c$ (days)& 
$T_{oc}$ (BKJD)& 
$\sqrt{e_c}\cos{\omega_c}$& 
$\sqrt{e_c}\sin{\omega_c}$& 
$i_c$ (deg)& 
$\Omega_c$ (deg)& 
$M_c (M_{Jup})$& 
$R_c (R_\star)$\\
\hline
 & 
4.00502692& 
3371.13169& 
-0.1995415& 
0.19396097& 
87.5118793& 
1.48106165& 
0.02360255& 
0.0599834
\\
 & 
4.00483110& 
3371.13112& 
-0.2054754& 
0.22162647& 
87.4948920& 
2.44051956& 
0.02378850& 
0.061674
\\
 & 
4.00498325& 
3371.13186& 
-0.1783291& 
0.17854464& 
87.5673562& 
1.99859017& 
0.02360026& 
0.0608582\\
 &
... &
... &
... &
... &
... &
... &
... &
...
\enddata
\end{deluxetable*}

The highest-likelihood set of lightcurve parameters from those discussed in Section~\ref{sec:lcresults} is shown in Table~\ref{tab:achbestfit}. The initial conditions of the best-fitting dynamical model from Section~\ref{sec:ttvresults} are in table~\ref{tab:bestfit}.

At times, our AIMCMC process for fitting the photometric data strongly constrained $e \cos{\omega}$. This was typically a constraint that it be close to $+0.3$ for planet b, but it was also sometimes constrained to be close to $-0.3$ and sometimes constrained for planet c. Because the effect of $e \cos{\omega}$ on transit duration is weak and symmetric, we believed this to be due to an error. The cause, we found, was a small shift in the true anomaly of transits that becomes noticeable only for inclined, eccentric orbits. We can define an axis perpendicular to the line of sight and parallel to the plane of the orbit. If we call this the x-axis, then one might expect the center of a transit to occur when the planet is at $x=0$, or
\begin{equation}
    f = \frac{\pi}{2} - \omega
\end{equation}
\noindent
However, in the sky-plane, an orbit that is both inclined and eccentric is not necessarily parallel to the x-axis at its closest approach to the center of the star. It can be shown that the true anomaly at which the center of the transit appears to be from photometry (the deepest point of the light curve, or halfway between ingress and egress for an orbit that isn't highly eccentric) is instead

\begin{equation}
    \label{eqn:del_f}
    f = \frac{\pi}{2} - \omega - \frac{e \cos{\omega} \cos^2(i)}{1+e \sin{\omega}}
\end{equation}
\noindent
to leading order in $e$ and $\cos{i}$. This leads to a timing offset of

\begin{equation}
    \label{eqn:del_t}
    \begin{split}
        \Delta t & = \Delta f \frac{(1-e^2)^{3/2}}{(1+e \sin{\omega})^2} \frac{T}{2 \pi}\\
        & = \frac{-e \cos{\omega}\cos^2(i)(1-e^2)^{3/2}}{(1+e\sin{\omega})^3}\frac{T}{2 \pi}
    \end{split}
\end{equation}

This meant that modeled lightcurves with values of $e \cos{\omega}$ far from zero effectively had their centers shifted forwards or backwards in time. Because of this, $e \cos{\omega}$ effectively served as an undesired degree of freedom offsetting transit times on a campaign-by-campaign basis. Because planet b has an inclination close to $90^\circ$, this offset would generally have been only about 10 seconds for $|e \cos{\omega}| = |e \sin{\omega}| = 0.2$ even in C16/18. For planet c, though, the offset for the same eccentricities would have reached beyond 30 seconds. Because this effect is small, it takes a large $e \cos{\omega}$ offset to create a given timing offset, especially for planet c. To obtain the results presented in this paper, we used \texttt{batman} models offset by a time equal and opposite to that listed in Equation~\ref{eqn:del_t} in order to avoid this effect and its erroneous biasing of $e \cos{\omega}$.

\clearpage
\begin{table}[tb]
    \caption{Expected future transit times, for planet b.}
    \centering % centering table
    \begin{tabular}{cccc} \toprule
         Planet & Transit Number & Transit time & Uncertainty\\
          & & (BKJD) & (minutes)\\
          \hline
         b   &  38 &    3472.7760 &   0.5 \\
 &  39 &    3475.4451 &   0.5 \\
 &  40 &    3478.1139 &   0.5 \\
 &  41 &    3480.7802 &   0.5 \\
 &  ... &   ...  & ... \\
 & 568 &    4880.9795 &  16.8 \\
 & 569 &    4883.6502 &  16.4 \\
 & 570 &    4886.3208 &  16.2 \\
 & 571 &    4888.9916 &  15.9 \\
 & 572 &    4891.6610 &  15.4 \\
 & 573 &    4894.3304 &  14.9 \\
 & 574 &    4897.0000 &  14.4 \\
 & 575 &    4899.6673 &  13.5 \\
 & 576 &    4902.3347 &  12.8 \\
 & 577 &    4905.0023 &  12.1 \\
 & 578 &    4907.6667 &  11.0 \\
 & 579 &    4910.3313 &  10.2 \\
 & 580 &    4912.9961 &   9.3 \\
 & 581 &    4915.6567 &   8.4 \\
 & 582 &    4918.3177 &   7.7 \\
 & 583 &    4920.9788 &   7.2 \\
 & 584 &    4923.6353 &   7.1 \\
 & 585 &    4926.2921 &   7.4 \\
 & 586 &    4928.9490 &   8.1 \\
 & 587 &    4931.6015 &   8.9 \\
 & 588 &    4934.2542 &   9.9 \\
 & 589 &    4936.9070 &  11.1 \\
 & 590 &    4939.5559 &  12.0 \\
 & 591 &    4942.2052 &  13.1 \\
 & 592 &    4944.8544 &  14.3 \\
 & 593 &    4947.5008 &  15.0 \\
 & 594 &    4950.1473 &  15.9 \\
 & 595 &    4952.7937 &  16.8 \\
 & 596 &    4955.4384 &  17.4 \\
 & 597 &    4958.0829 &  18.0 \\
 & 598 &    4960.7276 &  18.6 \\
 & 599 &    4963.3711 &  18.9 \\
 & 600 &    4966.0145 &  19.3 \\
 & 601 &    4968.6580 &  19.7 \\
 & 602 &    4971.3011 &  19.8 \\
 & 603 &    4973.9439 &  20.0 \\
 & 604 &    4976.5871 &  20.1 \\
 & 605 &    4979.2302 &  19.9 \\
 & 606 &    4981.8730 &  19.9 \\
 & 607 &    4984.5165 &  19.7 \\
 & 608 &    4987.1603 &  19.4 \\
 & 609 &    4989.8035 &  19.1 \\
 & 610 &    4992.4481 &  18.6 \\
           \hline
    \end{tabular}
    \label{tab:futureb}
\end{table}

 \begin{table}[tb]
    \caption{Expected future transit times, for planet c.}
    \centering % centering table
    \begin{tabular}{cccc} \toprule
         Planet & Transit Number & Transit time & Uncertainty\\
          & & (BKJD) & (minutes)\\
          \hline
         c   &  25 &    3470.5692 &   0.6 \\
 &  26 &    3474.5371 &   0.6 \\
 &  27 &    3478.5058 &   0.6 \\
 &  28 &    3482.4772 &   0.6 \\
 &  29 &    3486.4493 &   0.6 \\
 & ... & ... & ... \\
  & 374 &    4861.7518 &  16.8 \\
 & 375 &    4865.7170 &  17.2 \\
 & 376 &    4869.6814 &  17.5 \\
 & 377 &    4873.6461 &  17.6 \\
 & 378 &    4877.6103 &  17.5 \\
 & 379 &    4881.5752 &  17.4 \\
 & 380 &    4885.5402 &  17.0 \\
 & 381 &    4889.5062 &  16.5 \\
 & 382 &    4893.4730 &  15.7 \\
 & 383 &    4897.4409 &  14.9 \\
 & 384 &    4901.4109 &  13.7 \\
 & 385 &    4905.3817 &  12.5 \\
 & 386 &    4909.3560 &  11.0 \\
 & 387 &    4913.3311 &   9.5 \\
 & 388 &    4917.3111 &   8.0 \\
 & 389 &    4921.2916 &   7.0 \\
 & 390 &    4925.2779 &   6.7 \\
 & 391 &    4929.2648 &   7.3 \\
 & 392 &    4933.2571 &   8.5 \\
 & 393 &    4937.2503 &  10.2 \\
 & 394 &    4941.2479 &  11.7 \\
 & 395 &    4945.2465 &  13.4 \\
 & 396 &    4949.2481 &  14.6 \\
 & 397 &    4953.2508 &  15.9 \\
 & 398 &    4957.2552 &  16.8 \\
 & 399 &    4961.2606 &  17.7 \\
 & 400 &    4965.2668 &  18.3 \\
 & 401 &    4969.2738 &  18.8 \\
 & 402 &    4973.2808 &  19.0 \\
 & 403 &    4977.2884 &  19.1 \\
 & 404 &    4981.2954 &  19.0 \\
 & 405 &    4985.3023 &  18.6 \\
           \hline
    \end{tabular}
    \label{tab:futurec}
\end{table}

\begin{figure*}[!tbh]
  \begin{center}
    \includegraphics[width=0.96\textwidth, trim={0.0cm 0cm 0cm 0cm}, clip=true]{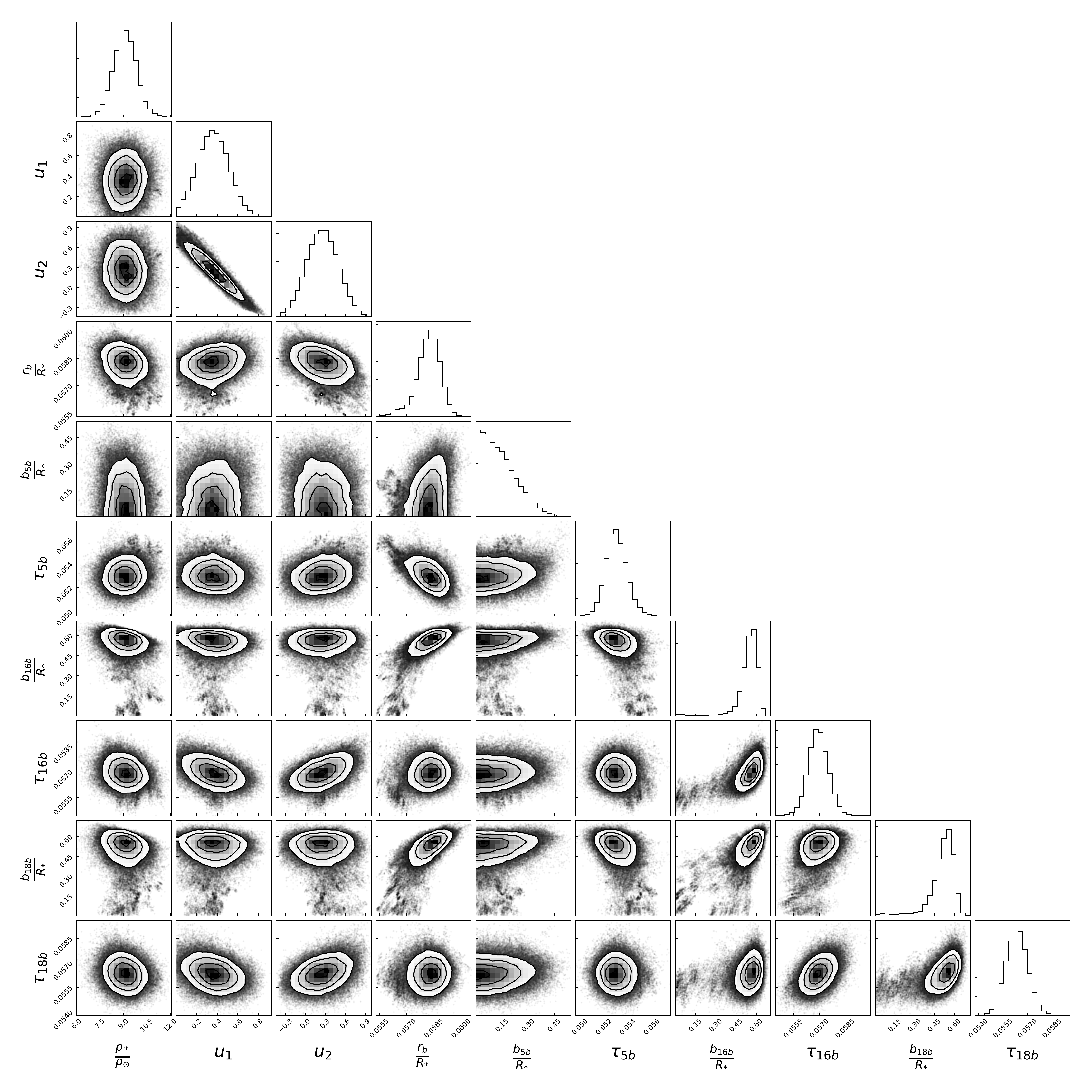}
   \end{center}
  \caption{Corner plot showing our posteriors on stellar density, LDPs, and features of planet b's transit after lightcurve fitting (but before dynamical fitting). Each diagonal plot shows a 1D histogram for samples of one parameter. Off-diagonal plots show two parameters at once, with contours replacing data points in regions of high sample density. The data in this plot represent 1/30th of the 4 million samples in our posterior set. $e\sin{\omega}$ and $e\cos{\omega}$ are only constrained through their combined influence on the more directly observable transit duration; thus, we show duration here rather than the eccentricity components. There is a noticeable tail of points with $b_{16b}$ and/or $b_{18b}$ reaching down to 0, many sigma from the most likely values. It is possible that a more extended burn-in period would have eliminated these tails. The number of points in one or both tails appears to be less than about 5\%, so we do no expect the tails to have a large effect on our median and uncertainty estimates of other parameters. Some correlations between parameters are evident, such as the strong negative correlation between $u_1$ and $u_2$ mentioned in Section~\ref{sec:lcresults}.}
  \label{fig:corner inner}
\end{figure*}

\begin{figure*}[!tbh]
  \begin{center}
    \includegraphics[width=0.96\textwidth, trim={0.0cm 0cm 0cm 0cm}, clip=true]{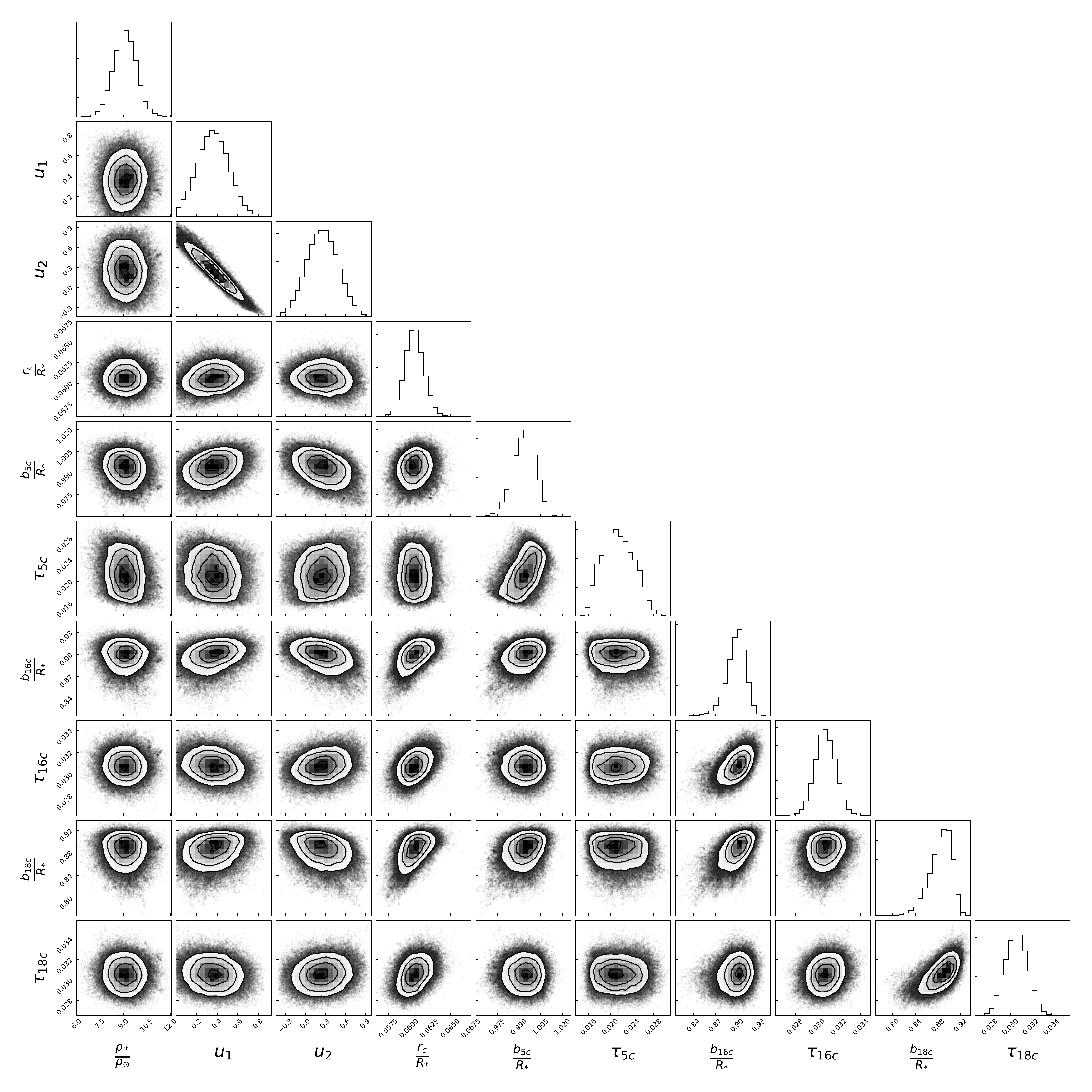}
   \end{center}
  \caption{Analogous to Figure~\ref{fig:corner inner}, except showing parameters of planet c rather than planet b. Planet c's impact parameter is more strongly constrained to large values, so the tails seen in Figure~\ref{fig:corner inner} are not present here.}
  \label{fig:corner outer}
\end{figure*}

\begin{figure*}[!tbh]
  \begin{center}
    \includegraphics[width=0.96\textwidth, trim={0.0cm 0cm 0cm 0cm}, clip=true]{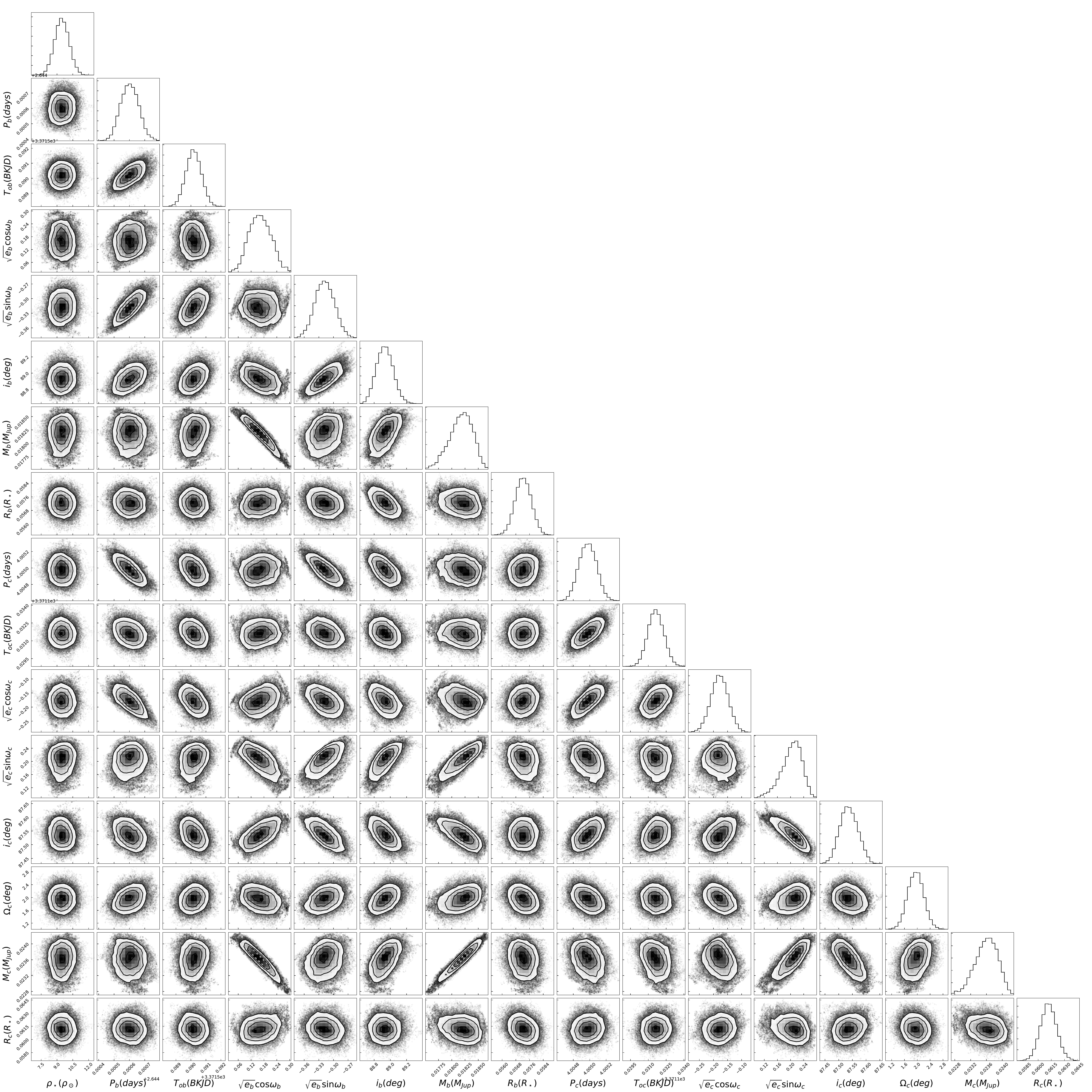}
   \end{center}
  \caption{ {\bf Parameter posteriors of dynamical model.} } 
  \label{fig:cornerTTV}
\end{figure*}

\end{document}